\documentclass[11pt,onecolumn]{article}

\usepackage{fullpage}

\usepackage{epsfig}
\usepackage{color}
\usepackage{graphicx}
\usepackage{amssymb,latexsym,amsmath}
\usepackage[latin9]{inputenc}
\usepackage{mathabx}
\usepackage{enumerate}
\usepackage{url}
\usepackage{pstricks}
\usepackage{relsize}
\usepackage{amsfonts,epsfig,graphicx}
\usepackage{amsmath,amssymb,amsthm}

\usepackage{multirow}
\usepackage{psfrag}

\makeatletter \long\def\@makecaption#1#2{
        \vskip 0.8ex
        \setbox\@tempboxa\hbox{\small {\bf #1:} #2}
        \parindent 1.5em  
        \dimen0=\hsize
        \advance\dimen0 by -3em
        \ifdim \wd\@tempboxa >\dimen0
                \hbox to \hsize{
                        \parindent 0em
                        \hfil
                        \parbox{\dimen0}{\def\baselinestretch{0.96}\small
                                {\bf #1.} #2
                                }
                        \hfil}
        \else \hbox to \hsize{\hfil \box\@tempboxa \hfil}
        \fi
        }
\makeatother

\newtheorem{theorem}{Theorem}
\newtheorem{lemma}{Lemma}

\theoremstyle{plain}

\theoremstyle{definition}

\newtheorem{asss}{Assumption}

\newcommand{\MySD}{MbSD }

\newcommand{\widgraph}[2]{\includegraphics[keepaspectratio,width=#1]{#2}}
\newcommand{\matsnorm}[2]{|\!|\!| #1 |\!|\!|_{{#2}}}
\newcommand{\vnorm}[2]{\ensuremath{\|#1\|_#2}}

\newcommand{\order}{\ensuremath{\mathcal{O}}}

\newcommand{\vecineq}{\ensuremath{\preceq}}
\newcommand{\real}{\ensuremath{\mathbb{R}}}
\newcommand{\defn}{\ensuremath{:=}}

\newcommand{\Prob}{\mathbb{P}} 
\newcommand{\Expt}{\mathbb{E}}
\newcommand{\var} {\operatorname{var}}
\newcommand{\modtwo}{\operatorname{\stackrel{2}{\equiv}}}

\newcommand{\paritymat}{\ensuremath{H}}
\newcommand{\graph}{\mathcal{G}}
\newcommand{\varset}{\mathcal{V}}
\newcommand{\chkset}{\mathcal{C}}
\newcommand{\edge}{\mathcal{E}}

\newcommand{\neig}{\ensuremath{\mathcal{N}}}
\newcommand{\compneigchk}{\ensuremath{\svar \in \neig(\fchk) \setminus
    \{\fvar\}}} 
\newcommand{\compneigvar}{\ensuremath{\schk \in
    \neig(\fvar) \setminus \{\fchk\}}}

\newcommand{\binsumopt}{\ensuremath{\bigoplus}}
\newcommand{\binequopt}{\ensuremath{\bigocirc}} 

\newcommand{\numvar}{\ensuremath{n}}
\newcommand{\numchk}{\ensuremath{m}}

\newcommand{\fvar}{\ensuremath{i}}
\newcommand{\svar}{\ensuremath{j}}
\newcommand{\fchk}{\ensuremath{a}}
\newcommand{\schk}{\ensuremath{b}}

\newcommand{\fvarfchk}{\ensuremath{\fvar\to\fchk}} 
\newcommand{\fchkfvar}{\ensuremath{\fchk\to\fvar}} 
\newcommand{\svarfchk}{\ensuremath{\svar\to\fchk}} 
\newcommand{\fchksvar}{\ensuremath{\fchk\to\svar}} 
\newcommand{\schkfvar}{\ensuremath{\schk\to\fvar}} 
\newcommand{\schksvar}{\ensuremath{\schk\to\svar}} 
\newcommand{\genedge}[2]{\ensuremath{#1\to#2}}

\newcommand{\chkdeg}{\ensuremath{d_c}}
\newcommand{\vardeg}{\ensuremath{d_v}}

\newcommand{\trabit}{\ensuremath{x}}
\newcommand{\recbit}{\ensuremath{y}}

\newcommand{\bpmes}{\ensuremath{\alpha}}
\newcommand{\bpmarg}{\ensuremath{\mu}}

\newcommand{\binsdmes}{\ensuremath{Z}}
\newcommand{\auxsdmes}{\ensuremath{W}}
\newcommand{\binsdmarg}{\ensuremath{U}}
\newcommand{\sdmarg}{\ensuremath{\hat{\eta}}}
\newcommand{\avesdmes}{\ensuremath{\beta}}

\newcommand{\tranzo}{\ensuremath{f_{\fvar}^{\Time}}}
\newcommand{\tranoz}{\ensuremath{g_{\fvar}^{\Time}}}
\newcommand{\tranzoEdge}{\ensuremath{f_{\fvarfchk}^{\Time}}}
\newcommand{\tranozEdge}{\ensuremath{g_{\fvarfchk}^{\Time}}}

\newcommand{\ffun}[3]{\ensuremath{f_{#1}^{#2}(#3)}}
\newcommand{\gfun}[3]{\ensuremath{g_{#1}^{#2}(#3)}}
\newcommand{\fsim}{\ensuremath{f}}
\newcommand{\gsim}{\ensuremath{g}}

\newcommand{\dimn}{\ensuremath{k}}
\newcommand{\Time}{\ensuremath{t}}
\newcommand{\ftime}{\ensuremath{T}}
\newcommand{\accu}{\ensuremath{\delta}}
\newcommand{\eps}{\ensuremath{\epsilon}}

\newcommand{\lbnd}{\ensuremath{c^{\ast}}}
\newcommand{\lips}{\ensuremath{M}}
\newcommand{\error}{\ensuremath{e}}
\newcommand{\diam}{\ensuremath{L}}

\newcommand{\numedge}{\ensuremath{r}}
\newcommand{\mat}{\ensuremath{A}}
\newcommand{\onevec}{\ensuremath{\vec{1}}}
\newcommand{\zerovec}{\ensuremath{\vec{0}}}

\begin{document}

\begin{center}

{\bf{\LARGE{ A Novel Stochastic Decoding of LDPC Codes with
      Quantitative Guarantees}}}

\vspace*{.2in}


{\large{
\begin{tabular}{ccc}
Nima Noorshams & & Aravind Iyengar\\ 
{\texttt{nnoorshams@cal.berkeley.edu}} & &
{\texttt{ariyenga@qti.qualcomm.com}}
\end{tabular}
}}\\
\vspace*{.1in}
Qualcomm Research Silicon Valley\\
Santa Clara, CA, USA

\vspace*{.1in}
\today

\end{center}

\begin{abstract}

Low-density parity-check codes, a class of capacity-approaching linear
codes, are particularly recognized for their efficient decoding
scheme. The decoding scheme, known as the sum-product, is an iterative
algorithm consisting of passing messages between variable and check
nodes of the factor graph. The sum-product algorithm is fully
parallelizable, owing to the fact that all messages can be update
concurrently. However, since it requires extensive number of highly
interconnected wires, the fully-parallel implementation of the
sum-product on chips is exceedingly challenging. Stochastic decoding
algorithms, which exchange binary messages, are of great interest for
mitigating this challenge and have been the focus of extensive
research over the past decade. They significantly reduce the required
wiring and computational complexity of the message-passing
algorithm. Even though stochastic decoders have been shown extremely
effective in practice, the theoretical aspect and understanding of
such algorithms remains limited at large. Our main objective in this
paper is to address this issue. We first propose a novel algorithm
referred to as the Markov based stochastic decoding. Then, we provide
concrete quantitative guarantees on its performance for
tree-structured as well as general factor graphs. More specifically,
we provide upper-bounds on the first and second moments of the error,
illustrating that the proposed algorithm is an asymptotically
consistent estimate of the sum-product algorithm. We also validate our
theoretical predictions with experimental results, showing we achieve
comparable performance to other practical stochastic decoders.

\end{abstract}


\section{Introduction}
\label{SecIntro}

Sparse graph codes, most notably low-density parity-check (LDPC), have
been adopted by the latest wireless communication
standards~\cite{TTSI,IEEE802.16e,Ghn,IEEEP802.3an}. They are known to
approach the channel capacity~\cite{RicUrb,MacKay99,
  Richardson01a,Richardson01b}. What makes them even more appealing
for practical purposes is their simple decoding
scheme~\cite{AjiMce00,KasEtal01}. More specifically, LDPC codes are
decoded via a message-passing algorithm called the sum-product
(SP). It is an iterative algorithm consisting of passing messages
between variable and check nodes in the factor
graph~\cite{AjiMce00,KasEtal01}. The fact that all messages in the SP
algorithm can be updated concurrently, makes the fully-parallel
implementation---where the factor graph is directly mapped onto the
chip---most efficient. However, due to complex and seemingly random
connections between check and variable nodes in the factor graph,
fully-parallel implementation of the SP is challenging. The wiring
complexity has a big impact on the circuit area and power
consumption. Also longer, more inter-connected wires can create more
parasitic capacitance and limit the clock rate.

Various solutions have been suggested by researchers in order to
reduce the implementation complexity of the fully-parallel SP
algorithm. Analog circuits have been designed for short LDPC
codes~\cite{HematiEtal06, WinsteadEtal06}. Bit serial algorithms,
where messages are transmitted serially over single wires, have been
proposed~\cite{DarabihaEtal06,DarabihaEtal08,BrandonEtal08,CushonEtal10}.
Splitting row-modules by partitioning check node operations has been
shown to provide substantial gains in the required area and power
efficiency~\cite{MohBaa06, MohseninEtal10}. In another prominent line
of work, researchers have proposed various stochastic decoding
algorithms~\cite{GauRap03, WinsteadEtal05, TehraniEtal06,
  TehraniEtal10, NaderiEtal11, KuolunEtal11, PrimeauEtal13,
  SarkisEtal13}. They are all based on stochastic representation of
the SP messages. More precisely, messages are encoded via Bernoulli
sequences with correct marginal probabilities. As a result, the
structure of check and variable nodes are substantially simplified and
the wiring complexity is significantly reduced. (Benefits of such
decoders are discussed in more details in
Section~\ref{SubSecImplementation}.) Stochastic message-passing have
also been used in other contexts, among which are distributed convex
optimization and learning~\cite{JuditskyEtal09, HazanEtal07},
efficient belief propagation algorithms~\cite{NooWai13b, NooWai13a},
and efficient learning of distributions~\cite{SarJav11}.

Although experimental results have proved stochastic decoding
extremely beneficial, to date mathematical understanding of such
decoders are very limited and largely missing from the
literature. Since the output of stochastic decoders are random by
construction, it is natural to ask the following questions: how does
the stochastic decoder behave on average? can it be tuned to approach
the performance of SP and if so how fast? is the average performance
typical or do we have a measure of concentration around average? The
main contribution of this paper is answering these questions by
providing \emph{theoretical analysis} for stochastic decoders. To that
end, we propose a novel algorithm, referred to as Markov based
stochastic decoding (MbSD), which is amenable to theoretical analysis.
We provide quantitative bounds on the first and second moments of the
error in terms of the underlying parameters for tree-structured (cycle
free) as well as general factor graphs, showing that the performance
of \MySD converges to that of SP.

The remainder of this paper is organized as follows. We begin in
Section~\ref{SecProbState} with some background on factor graph
representation of LDPC codes, the sum-product algorithm, and
stochastic decoding. In Section~\ref{SecResults}, we turn to our main
results by introducing the \MySD algorithm followed by some discussion
on its hardware implementation and statements of our main theoretical
results (Theorems~\ref{ThmTree},
and~\ref{ThmMain}). Section~\ref{SecProof} is devoted to proofs, with
some technical aspects deferred to appendices. Finally in
Section~\ref{SecSimulations}, we provide some experimental results,
confirming our theoretical predictions.


\section{Background and Problem Setup}
\label{SecProbState}

In this section, we setup the problem and provide the necessary
background.

\subsection{Factor Graph Representation of LDPC Codes}
\label{SubSecFactorGraph}

\begin{figure}
\begin{center}
  \widgraph{.55\textwidth}{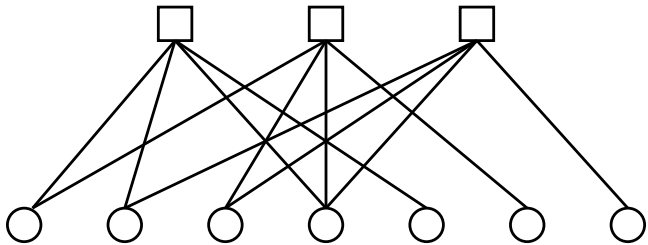}
\end{center}
\caption{Factor graph of the Hamming code. Variable nodes are
  represented by circles, whereas check nodes are represented by
  squares. }\label{FactorGraph}
\end{figure}

A low-density parity-check code is a linear error-correcting code,
satisfying a number of parity check constraints. These constraints are
encoded by a sparse parity-check matrix $\paritymat \in \{0,
1\}^{\numchk\times\numvar}$. More specifically, a binary sequence
$\trabit \in \{0, 1\}^{\numvar}$ is a valid codeword if and only if
$\paritymat\trabit \modtwo 0$, where all operations are module
two~\cite{RicUrb}. A popular approach for modeling LDPC codes is via
the notion of factor graphs~\cite{KasEtal01}. A factor graph
representing an LDPC code with the parity-check matrix $\paritymat$ is
a bipartite graph $\graph = (\varset, \chkset, \edge)$, consisting of
a set of variable nodes $\varset \defn \{1, 2, \ldots, \numvar\}$, a
set of check nodes $\chkset \defn \{1, 2, \ldots, \numchk\}$, and a
set of edges connecting variable and check nodes $\edge \defn
\{(\fvar, \fchk) \mid \fvar \in \varset, \fchk \in \chkset, \,
\text{and} \, \paritymat(\fchk, \fvar) = 1\}$. (In this paper, we use
letters $\fvar, \svar, \ldots$, and $\fchk, \schk, \ldots$ to denote
variable and check nodes respectively.) A typical factor graph
representing an LDPC code (the Hamming code) is illustrated in
Figure~\ref{FactorGraph}.

\subsection{The Sum-Product Algorithm}
\label{SubSecSP}

Suppose a transmitter sends the codeword $\trabit$ to a receiver over
a memory-less, noisy communication channel. Some channel models that
are commonly used in practice include the additive white Gaussian
noise (AWGN), the binary symmetric channel, and the binary erasure
channel. Having received the impaired signal $\recbit$, the receiver
attempts to recover the original signal by finding either the global
maximum aposteriori (MAP) estimate $\widehat{\trabit} \; = \;
\arg\max_{\paritymat\trabit = 0} \, \Prob(\trabit \mid \recbit)$, or
the bit-wise MAP estimates \mbox{$\widehat{\trabit}_i \; = \;
  \arg\max_{\paritymat\trabit = 0} \, \Prob(\trabit_i \mid \recbit)$},
for $i = 1, 2, \ldots, \numvar$.

Without exploiting the underlying structure of the code, or
equivalently its factor graph, the MAP estimation is intractable and
requires an exponential number of operations in the code
length. However, this problem can be circumvented using an algorithm
called the sum-product (SP), also known as the belief propagation
algorithm. The SP is an iterative algorithm consisting of passing
messages, in the form of probability distributions, between nodes of
the factor graph~\cite{AjiMce00,KasEtal01}. It is known to converge to
the correct bit-wise MAP estimates for cycle-free factor graphs;
however, on loopy graphs, which includes almost all practical LDPC
codes, such a guarantee no longer exists. Nonetheless, the SP
algorithm has been shown to be extremely accurate and effective in
practice~\cite{MacKay99, RicUrb}. 

\begin{figure}
\begin{center}
\begin{tabular}{ccc}
\psfrag{*i*}{$\fvar$}
\psfrag{*j1*}{$\svar_1$}
\psfrag{*j2*}{$\svar_2$}
\psfrag{*j3*}{$\svar_3$}
\psfrag{*a*}{$\fchk$}
\widgraph{.42\textwidth}{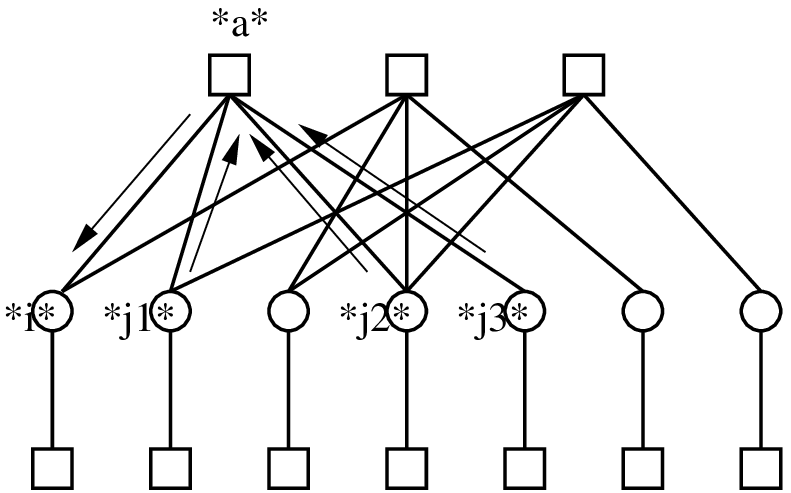} & & 
\psfrag{*i*}{$\fvar$}
\psfrag{*b*}{$\schk$}
\psfrag{*a*}{$\fchk$}
\widgraph{.42\textwidth}{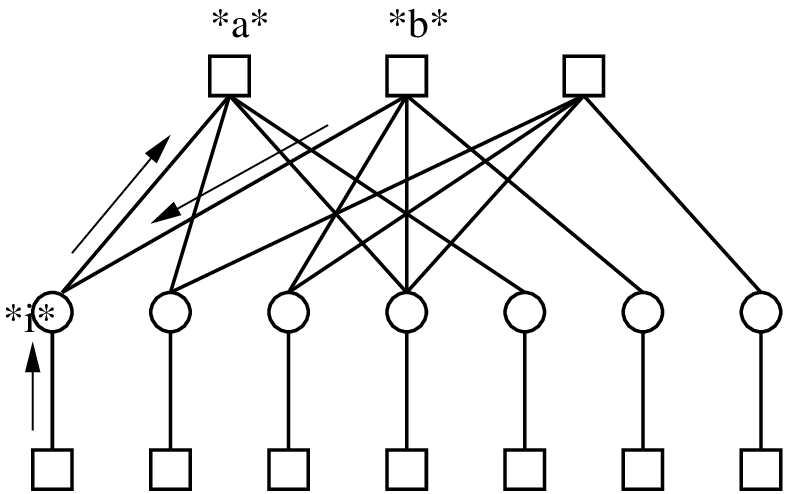}\\
(a) & & (b)
\end{tabular}
\end{center}
\caption{Graphical representation of message-passing on factor graphs
  (a) check to variable node (b) variable to check node.}
\label{FigMessageFlow}
\end{figure}

We now turn to the description of the SP algorithm. For every variable
node $\fvar\in\varset$ let $\neig(\fvar) \defn \{\fchk \mid (\fvar,
\fchk) \in \edge\}$ denote the set of its neighboring check
nodes. Similarly define $\neig(\fchk) \defn \{\fvar \mid (\fvar,
\fchk) \in \edge\}$, the set of neighboring variable nodes for every
check node $\fchk\in\chkset$. The SP algorithm allocates two messages
to every edge $(\fvar, \fchk)\in\edge$, one for each direction. At
each iteration $\Time = 0, 1, \ldots$, every variable node
$\fvar\in\varset$ (check node $\fchk\in\chkset$), calculates a message
$0 < \bpmes_{\fvarfchk}^{\Time+1} < 1$ (message $0 <
\bpmes_{\fchkfvar}^{\Time+1} < 1$) and transmit it to its neighboring
check node $\fchk\in\neig(\fvar)$ (variable node
$\fvar\in\neig(\fchk)$). In updating the messages, every variable node
takes into account the incoming messages from its neighboring check
nodes as well as the information from the channel, namely
$\bpmes_{\fvar} \defn \Prob(\trabit_{\fvar} = 1 \mid
\recbit_{\fvar})$. With this notation at hand, the description of the
SP algorithm is as follows: initialize messages from variable to check
nodes, $\bpmes_{\fvarfchk}^{0} = \bpmes_{\fvar}$, and update messages
for each edge $(\fvar, \fchk)\in\edge$ and iteration $\Time = 0, 1,
\ldots$ according to
\begin{align}\label{EqnSPchk}
\bpmes_{\fchkfvar}^{\Time} \; = \; \frac{1}{2} \, - \, \frac{1}{2}
\prod_{\compneigchk}(1 - 2\:\bpmes_{\svarfchk}^{\Time}), 
\end{align}
and 
\begin{align}\label{EqnSPvar}
\bpmes_{\fvarfchk}^{\Time+1} \; = \;
\frac{\bpmes_{\fvar}\prod_{\compneigvar}\bpmes_{\schkfvar}^{\Time}}
     {\bpmes_{\fvar}\prod_{\compneigvar}\bpmes_{\schkfvar}^{\Time} \,
       + \,
       (1-\bpmes_{\fvar})\prod_{\compneigvar}(1-\bpmes_{\schkfvar}^{\Time})}.
\end{align}
Information flow on a factor graph is shown in
Figure~\ref{FigMessageFlow}. Upon receiving all the incoming messages,
variable node $\fvar\in\varset$ update its marginal probability
\begin{align*}
\bpmarg_\fvar^{\Time+1} \; = \;
\frac{\bpmes_{\fvar}\prod_{\schk\in\neig(\fvar)}\bpmes_{\schkfvar}^{\Time}}
     {\bpmes_{\fvar}\prod_{\schk\in\neig(\fvar)}\bpmes_{\schkfvar}^{\Time}
       \, + \,
       (1-\bpmes_{\fvar})\prod_{\schk\in\neig(\fvar)}(1-\bpmes_{\schkfvar}^{\Time})}.
\end{align*}
Accordingly, the receiver estimates the $\fvar$-th bit by
$\widehat{\trabit}_i^{\Time+1} = \mathbb{I}(\bpmarg_\fvar^{\Time+1} >
0.5)$, where $\mathbb{I}(\cdot)$ is the indicator function. It should
also be mentioned that in practice, in order to reduce the
quantization error, log-likelihood ratios are mostly used as
messages. Moreover, to further simplify the SP algorithm, the check
node operation~\eqref{EqnSPchk} is approximated. The resultant is
known as the Min-Sum algorithm~\cite{FossorierEtal99}.

\subsection{Stochastic Decoding of LDPC Codes}
\label{SubSecSD}

Stochastic computation in the context of LDPC decoding was first
introduced in 2003 by Gaudet and Rapley~\cite{GauRap03}. Ever since,
much research has been conducted in this field and numerous stochastic
decoders have been proposed~\cite{WinsteadEtal05, TehraniEtal06,
  TehraniEtal10, NaderiEtal11, PrimeauEtal13, SarkisEtal13}. For
instance, Tehrani et al.~\cite{TehraniEtal06} introduced and exploited
the notions of edge memory and noise dependent scaling in order to
make the stochastic decoding a viable method for long, practical, LDPC
codes. Estimating the probability distributions via a successive
relaxation method, Leduce-Primeau et al.~\cite{PrimeauEtal13} proposed
a scheme with improved decoding gain. More recently, Sarkis et
al.~\cite{SarkisEtal13} extended the stochastic decoding to the case
of non-binary LDPC codes.

The underlying structure of all these methods and most relevant to our
work, however, are the following: they all encode messages by
Bernoulli sequences, they all consist of `decoding cycles' which
should not be confused with SP iterations (roughly speaking, multiple
decoding cycles correspond to one SP iteration.), the check node
operation is the module-two sum (i.e. the message transmitted from a
check node to a variable node is equal to the module-two sum of the
incoming bits.), and finally the variable node operation is the
equality (i.e. the message transmitted from a variable node to a check
node is equal to one if all incoming bits are one, it is equal to zero
if all incoming bits are zero, and it is equal to the previous
decoding cycle's bit in case incoming messages do not agree.). The
intuition behind the stochastic variable and check node operations can
be obtained from the inspection of SP message updates~\eqref{EqnSPchk}
and~\eqref{EqnSPvar}. More specifically, suppose
$\binsdmes_{\svarfchk}$, for $\compneigchk$, are independent Bernoulli
random variables with distributions $\Prob(\binsdmes_{\svarfchk} = 1)
= \bpmes_{\svarfchk}^{\Time}$. Then $\bpmes_{\fchkfvar}^{\Time}$,
derived from equation~\eqref{EqnSPchk}, becomes the probability of
having odd number of ones in the sequence
$\{\binsdmes_{\svarfchk}\}_{\compneigchk}$ (see Lemma 1 in the
paper~\cite{Gallager62}). Therefore, the statistically consistent
estimate of the check to variable node message is the module-two
summation of the incoming bits. Similarly, to understand the
stochastic variable node operation, let $\binsdmes_{\fvar}$ and
$\binsdmes_{\schkfvar}$, for $\compneigvar$, be independent Bernoulli
random variables with probability distributions
$\Prob(\binsdmes_{\fvar} = 1) = \bpmes_{\fvar}$, and
$\Prob(\binsdmes_{\schkfvar} = 1) = \bpmes_{\schkfvar}^{\Time}$. Then
$\bpmes_{\fvarfchk}^{\Time+1}$, derived from
equation~\eqref{EqnSPvar}, becomes the probability of the event
$\{\binsdmes_{\fvar} = 1, \, \binsdmes_{\schkfvar} = 1, \,
\forall\compneigvar\}$, conditioned on the event $\{\binsdmes_{\fvar}
= \binsdmes_{\schkfvar}, \, \forall\compneigvar\}$, thus supporting
the intuition that one must transmit the common value from variable to
check nodes in case all incoming bits are equal.


\section{Algorithm and Main Results}
\label{SecResults}

In this section, we introduce the MbSD algorithm, discuss its hardware
design aspect, and state some theoretical guarantees regarding its
performance.

\subsection{The Proposed Stochastic Algorithm}
\label{SubSecAlgorithm}

The MbSD algorithm consists of passing messages between variable and
check nodes of the factor graph. These messages are
$2\dimn$-dimensional binary vectors, for a fixed $\dimn$ (design
parameter). However, variable and check node updates are element-wise,
bit operations. Before stating the algorithm we need to define some
notation.

Suppose $(\recbit_{1}, \recbit_{2}, \ldots, \recbit_{\numvar})$ is the
received codeword with the likelihood $\bpmes_{\fvar} =
\Prob(\trabit_{\fvar} = 1 | \recbit_{\fvar})$, for $\fvar = 1, 2,
\ldots, \numvar$. Our algorithm, involves messages from the channel to
variable nodes at every iteration $\Time=0,1,2,\ldots$. More
specifically, let $\binsdmes_{\fvar}^{\Time} \in \{0, 1\}^{2\dimn}$ be
the $2\dimn$-dimensional binary message from the channel to the
variable node $\fvar$ at time $\Time$, with independent and
identically distributed (i.i.d.)  entries
\begin{align*}
\Prob(\binsdmes_{\fvar}^{\Time}(\ell) = 1) \;=\; \bpmes_{\fvar},\quad
\text{for all $\ell = 1, 2, \ldots, 2\dimn$}.
\end{align*} 
Moreover, let $\binsdmes_{\fvarfchk}^{\Time} \in \{0, 1\}^{2\dimn}$
denote the $2\dimn$-dimensional binary message from the variable node
$\fvar$ to the check node $\fchk$ at time $\Time = 0, 1,
\ldots$. Similarly, let $\binsdmes_{\fchkfvar}^{\Time} \in \{0,
1\}^{2\dimn}$ be the message from the check node $\fchk$ to the
variable node $\fvar$ at time $\Time$.

We also need to define the element-wise, module-two summation operator
$\binsumopt$, as well as the ``equality'' operator
$\binequopt$. Suppose $X_1, X_2, \ldots, X_d$ are arbitrary
$2\dimn$-dimensional binary vectors. Then, the vector $Y =
\binsumopt_{i=1}^{d} X_i$ denotes the module-two summation of the
vectors $\{X_i\}_{i=1}^{d}$, if and only if
\begin{align*}
Y(\ell) \;\modtwo\; X_1(\ell) \,+\, X_2(\ell) \,+\, \ldots \,+\,
X_d(\ell),
\end{align*}
for all $\ell=1,2,\ldots,2\dimn$. Furthermore, by
$Y=\binequopt_{i=1}^{d} X_i$ we mean
\begin{align}\label{DefnEquOpt}
Y(\ell) \; = \; \left\{\begin{array}{lc} 1 & \text{if $X_i(\ell) = 1$,
  for all $i=1,2,\ldots,d$ } \\ 0 & \text{if $X_i(\ell) = 0$, for all
  $i=1,2,\ldots,d$ }\\ Y(\ell-1) & \text{otherwise}
\end{array}\right.,
\end{align}
for all $\ell = 1, 2, \ldots, 2\dimn$. Here, we assume $Y(0)$ is
either zero or one, equally likely.\\

\noindent Now, the precise description of the \MySD algorithm is as
follows:

\begin{enumerate}
\item Initialize messages from variable nodes to check nodes at time
  $\Time = 0$ by $\binsdmes_{\fvarfchk}^{0} = \binsdmes_{\fvar}^{0}$.

\item For iterations $\Time = 0, 1, 2, \ldots$, and every edge
  $(\fvar,\fchk)\in\edge$:
\begin{itemize}
\item[a)] Update messages from check nodes to variable nodes
\begin{align*}
\binsdmes_{\fchkfvar}^{\Time} \;\; = \binsumopt_{\compneigchk}
\binsdmes_{\svarfchk}^{\Time}.
\end{align*}
\item[b)] Update messages from variable nodes to check nodes by
  following these steps:
\begin{itemize}
\item compute the auxiliary variable
\begin{align*}
\auxsdmes_{\fvarfchk}^{\Time+1} \;\; = \binequopt_{\compneigvar}
\binsdmes_{\schkfvar}^{\Time} \; \binequopt \; \binsdmes_{\fvar}^{\Time+1}.
\end{align*}

\item update the message entries
  $\binsdmes_{\fvarfchk}^{\Time+1}(\ell)$, for $\ell = 1, 2, \ldots,
  2\dimn$, by drawing i.i.d. samples from the set
\begin{align*}
\{\auxsdmes_{\fvarfchk}^{\Time+1}(\dimn+1), \, 
\auxsdmes_{\fvarfchk}^{\Time+1}(\dimn+2), \, \ldots, \,
\auxsdmes_{\fvarfchk}^{\Time+1}(2\dimn)\}.
\end{align*}

\end{itemize}

\item[c)] Compute the binary vector $\binsdmarg_{\fvar}^{\Time+1} =
  \binequopt_{\fchk \in \neig(\fvar)} \binsdmes_{\fchkfvar}^{\Time}
  \binequopt \binsdmes_{\fvar}^{\Time+1}$ and update the marginal
  estimates according to
\begin{align}\label{EqnSDmarg}
\sdmarg_{\fvar}^{\Time+1} \; = \; \frac{1}{\dimn} \:
\sum_{\ell=\dimn+1}^{2\dimn} \binsdmarg_{\fvar}^{\Time+1}(\ell),
\end{align}
for all $\fvar=1, 2, \ldots, \numvar$.

\end{itemize}

\end{enumerate}

Few comments, regarding the interpretation of the algorithm, are worth
mentioning at this point. The check to variable node message update
(step (a)) is a statistically consistent estimate of the actual check
to variable BP message update. However, same can not be stated about
the variable to check node update (step (b)). As will be shown in
Section~\ref{SecProof}, the equality operator $\binequopt$ generates
Markov chains with desirable properties, thereby, justifying the
``Markov based stochastic decoding'' terminology. More specifically,
the sequence
$\{\auxsdmes_{\fvarfchk}^{\Time+1}(\ell)\}_{\ell=1}^{2\dimn}$ is a
Markov chain with the actual variable to check BP message as its
stationary distribution. Our objective in step (b) is to estimate this
stationary distribution. From basic Markov chain theory, we know that
the marginal distribution of a chain converges to its stationary
distribution. Therefore, for large enough $\dimn$, the empirical
distribution of the set $\{\auxsdmes_{\fvarfchk}^{\Time+1}(\dimn+1),
\auxsdmes_{\fvarfchk}^{\Time+1}(\dimn+2), \ldots,
\auxsdmes_{\fvarfchk}^{\Time+1}(2\dimn)\}$ becomes an accurate enough
estimate of the stationary distribution of the Markov chain
$\{\auxsdmes_{\fvarfchk}^{\Time+1}(\ell)\}_{\ell=1}^{2\dimn}$.


\subsection{Discussion on Hardware Implementation}
\label{SubSecImplementation}

The proposed decoding scheme enjoys all the benefits of traditional
stochastic decoders~\cite{GauRap03, TehraniEtal06}. Since messages
between variable and check nodes are binary, stochastic decoding
requires a substantially lower wiring complexity compared to
fully-parallel sum-product or min-sum implementations. Shorter wires
yield smaller circuit area, and smaller parasitic capacitance which in
turn lead to higher clock frequencies and less power
consumption. Another advantage of stochastic decoding algorithms is
the very simple structure of check and variable nodes. As a matter of
fact, check nodes can be carried out with simple XOR gates, and
variable nodes can be implemented using a combination of a random
number generator, a JK flip flop, and AND
gates~\cite{GauRap03}. Finally, a very beneficial property of
stochastic decoding is the fact that the check node operation (XOR) is
associative, i.e., can be partitioned arbitrarily without introducing
any additional error. Mohsenin et al.~\cite{MohseninEtal10},
illustrated that partitioning check nodes can provide significant
improvements (by a factor of four) in area, throughput, and energy
efficiency.

It should be noted that in this paper and for mathematical
convenience, the messages between check and variable nodes are
represented by binary vectors. However, to implement the \MySD
algorithm, there is no need to buffer all these vectors. We only need
to count the number of ones between bits $\dimn+1$, and $2\dimn$ in
each bulk, which can be accomplished by a simple counter. In that
respect, \MySD has a great advantage compared to the algorithm
proposed by Tehrani et al.~\cite{TehraniEtal06}, which requires
buffering a substantial number of bits on each edge (edge
memories). As will be discussed in Section~\ref{SecSimulations}, our
algorithm has a superior bit error rate performance compared
to~\cite{TehraniEtal06}, while maintaining the same order of maximum
number of clocks, thereby achieving comparable if not better
throughput. Moreover, \MySD is equipped with concrete theoretical
guarantees, the subject to which we now turn.


\subsection{Main Theoretical Results}
\label{SubSecTheory}

Our results concern both cases of tree-structured (cycle free) as well
as general factor graphs. Since factor graphs of randomly generated
LDPC codes are locally tree-like~\cite{RicUrb}, understanding the
behavior of every decoding algorithm (stochastic as well as
deterministic) on trees is of paramount importance. To that end, we
first state some quantitative guarantees regarding the performance of
the proposed stochastic decoder on tree-structured factor graphs.

Recalling the fact that there exists a unique path between every two
variable nodes in a tree, we denote the largest path (also known as
the graph diameter) by $\diam$. Moreover, we know that estimates
generated by the SP algorithm on a tree converge to true marginals
after $\diam$ iterations~\cite{AjiMce00}, i.e., denoting true
marginals by $\{\bpmarg_{\fvar}^{\ast}\}_{\fvar=1}^{\numvar}$, we have
$\bpmarg_{\fvar}^{\Time} = \bpmarg_{\fvar}^{\ast}$, for all $\fvar =
1, 2, \ldots, \numvar$, and $\Time \ge \diam$.

\begin{theorem}[Trees]\label{ThmTree}
Consider the sequence of marginals
$\{\sdmarg_{\fvar}^{\Time+1}\}_{\Time=0}^{\infty}$, $\fvar = 1, 2,
\ldots, \numvar$, generated by the \MySD algorithm on a
tree-structured factor graph. Then for arbitrarily small but fixed
parameter $\accu$, and sufficiently large $\dimn = \dimn(\accu,
\graph)$ we have:
\begin{enumerate}[(a)]

\item The expected stochastic marginals become arbitrarily close to
  the true marginals, i.e.,
\begin{align*}
\max_{1\le\fvar\le\numvar}
\;\big|\Expt\big[\sdmarg_{\fvar}^{\Time}\big]\, - \,
\bpmarg_{\fvar}^{\ast}\big| \; \le \; \accu,
\end{align*}
for all $\Time \ge \diam$.
\item Furthermore, we have
\begin{align*}
\max_{1\le\fvar\le\numvar} \; \max_{\Time\ge0} \;
\var\big(\sdmarg_{\fvar}^{\Time}\big) \; = \;
\order\left(\frac{1}{\dimn}\right).
\end{align*}
\end{enumerate}

\end{theorem}

\paragraph{Remarks:} 
Theorem~\ref{ThmTree} provides quantitative bounds on the first and
second moments of the \MySD marginal estimates. Combining parts (a),
and (b), it can be easily observed that
\begin{align}\label{EqnConvRate}
\nonumber
\max_{1\le\fvar\le\numvar} \;\Expt\big[(\sdmarg_{\fvar}^{\Time}\, - \,
  \bpmarg_{\fvar}^{\ast})^2 \big] \; &\le \;
\max_{1\le\fvar\le\numvar}
\;\big|\Expt\big[\sdmarg_{\fvar}^{\Time}\big]\, - \,
\bpmarg_{\fvar}^{\ast}\big|^2 \,+\, \max_{1\le\fvar\le\numvar} \;
\var\big(\sdmarg_{\fvar}^{\Time}\big)\\ \; & \le \; \accu^2 \, + \,
\order\left(\frac{1}{\dimn}\right),
\end{align}
for all $\Time \ge \diam$. Therefore, as $\dimn\to\infty$
($\accu\to0$), the sequence of estimates $\sdmarg_{\fvar}^{\diam}$
(ranging over $\dimn$) converges to the true marginal
$\bpmarg_{\fvar}^{\ast}$ in the $\mathcal{L}^2$ sense. The rate of
convergence, and its dependence on the underlying parameters, is fully
characterized in expression~\eqref{EqnWhatIsK}. It is directly a
function of the accuracy, and the factor graph structure (diameter,
node degrees, etc.), and indirectly (through Lipschitz constants,
etc.) a function of the signal to noise ratio (SNR).\\

We now turn to the statement of results for LDPC codes with general
(loopy) factor graphs. Unlike tree-structured graphs, the existence
and uniqueness of the SP fixed points on general graphs is not
guaranteed, nor is the convergence of SP algorithm to such fixed
points. Therefore, we have to make the assumption that the LDPC code
of interest is well behaved. More precisely, we make the following
assumptions:

\begin{asss}\label{AsssConsistency}
Suppose the SP message updates are consistent, that is
$\bpmes_{\fvarfchk}^{\Time} \to \bpmes_{\fvarfchk}^{\ast}$, and
$\bpmes_{\fchkfvar}^{\Time} \to \bpmes_{\fchkfvar}^{\ast}$ as $\Time
\to \infty$ for all directed edges $(\fvarfchk)$, and
$(\fchkfvar)$. Equivalently, there exists a sequence
$\{\bpmarg_{\fvar}^{\ast}\}_{\fvar=1}^{\numvar}$ such that
$\bpmarg_{\fvar}^{\Time} \to \bpmarg_{\fvar}^{\ast}$, for all $\fvar =
1,2, \ldots,\numvar$.
\end{asss}
\noindent For an accuracy parameter $\accu > 0$, arbitrarily small, we
define the stopping time
\begin{align}\label{EqnDefnStop}
\ftime \; = \; \ftime(\accu) \; \defn \; \inf \big\{ \Time \: | \:
\max_{1\le\fvar\le\numvar} |\bpmarg_{\fvar}^{\Time} -
\bpmarg_{\fvar}^{\ast}| \, \le \, \accu \big\}.
\end{align}
According to assumption~\ref{AsssConsistency}, the stopping time
$\ftime$ is always finite.

\begin{theorem}[General factor graphs]\label{ThmMain}
Consider the marginals
$\{\sdmarg_{\fvar}^{\Time+1}\}_{\Time=0}^{\infty}$ generated by the
\MySD algorithm on an LDPC code that satisfies
Assumption~\ref{AsssConsistency}. Then for arbitrarily small but fixed
parameter $\accu$, and sufficiently large $\dimn = \dimn(\accu,
\ftime, \graph)$ we have:
\begin{enumerate}[(a)]

\item The expected stochastic marginals become arbitrarily close to
  the SP marginals, i.e.,
\begin{align*}
\big|\Expt\big[\sdmarg_{\fvar}^{\Time}\big]\, - \,
\bpmarg_{\fvar}^{\Time}\big| \; \le \; \accu,
\end{align*}
for all $\fvar = 1, 2, \ldots, \numvar$, and $\Time = 0, 1, \ldots, \ftime$.
\item Furthermore, we have
\begin{align*}
\max_{1\le\fvar\le\numvar} \; \max_{0\le\Time\le\ftime} \;
\var\big(\sdmarg_{\fvar}^{\Time}\big) \; = \;
\order\left(\frac{1}{\dimn}\right).
\end{align*}
\end{enumerate}

\end{theorem}

\paragraph{Remarks:}
Theorem~\ref{ThmMain}, in contrast to Theorem~\ref{ThmTree}, provides
quantitative bounds on the error over a finite horizon specified by
the stopping time~\eqref{EqnDefnStop}. After $\ftime = \ftime(\accu)$
iterations, the marginal estimates become arbitrarily close to the
true marginals on average; in particular, we have
\begin{align*}
\max_{0\le\fvar\le\numvar} |\Expt\big[\sdmarg_{\fvar}^{\ftime}\big] -
\bpmarg_{\fvar}^{\ast}| \; \le \; \max_{0\le\fvar\le\numvar}
|\Expt\big[\sdmarg_{\fvar}^{\ftime}\big] - \bpmarg_{\fvar}^{\ftime}| +
\max_{0\le\fvar\le\numvar} |\bpmarg_{\fvar}^{\ftime} -
\bpmarg_{\fvar}^{\ast}| \; \le \; 2\:\accu.
\end{align*}
Moreover, since $\var(\sdmarg_{\fvar}^{\ftime}) = \order(1/\dimn)$, as
$\dimn\to\infty$, the random variables
$\{\sdmarg_{\fvar}^{\ftime}\}_{\fvar=1}^{\numvar}$, become more and
more concentrated around their means. Specifically, a very crude
bound\footnote{Tightening this bound exploiting Chernoff inequality
  and concentration of measure~\cite{ChuLu06}, can be further
  explored.} using Chebyshev inequality~\cite{Grimmett} yields
\begin{align*}
\Prob\left(\max_{0\le\fvar\le\numvar} |\sdmarg_{\fvar}^{\ftime} -
\Expt\big[\sdmarg_{\fvar}^{\ftime}\big]| \ge \eps\right) \; \le \;
\numvar \: \max_{0\le\fvar\le\numvar} \:
\Prob\left(|\sdmarg_{\fvar}^{\ftime} -
\Expt\big[\sdmarg_{\fvar}^{\ftime}\big]| \ge \eps \right) \; = \;
\order\left(\frac{\numvar}{\dimn\:\eps^2}\right).
\end{align*}
Therefore, it is expected that the performance of the proposed
stochastic decoding converges to that of SP, as $\dimn\to\infty$.


\section{Proof of the Main Results}
\label{SecProof}

Conceptually, proofs of Theorems~\ref{ThmTree} and~\ref{ThmMain} are
very similar. Therefore, in this section, we only prove
Theorem~\ref{ThmTree} and highlight its important differences with
Theorem~\ref{ThmMain} in Appendix~\ref{AppProofThmMain}.

Poofs make use of basic probability and Markov chain theory. At a high
level, the argument consists of two parts: characterizing the expected
messages and controlling the error propagation in the factor graph. As
it turns out, the check node operations (module two summation
$\binsumopt$) are consistent on average, that is expected messages
from check to variable nodes are the same as SP messages. In contrast,
the variable node operations (equality operator $\binequopt$) are
asymptotically consistent (as $\dimn\to\infty$). Therefore, for a
finite message dimension $\dimn$, variable node operations introduce
error terms which become propagated throughout the factor graph. The
main challenge is to characterize and control these errors.

\subsection{Proof of Part $(a)$ of Theorem~\ref{ThmTree}}
\label{SubSecPartA}

We start by stating a lemma which plays a key role in the
sequel. Recall the definition of the equality operator $\binequopt$
from~\eqref{DefnEquOpt}.

\begin{lemma}\label{LemMarkovChain}
Suppose $\binsdmes_{i} = \{\binsdmes_{i}(\ell)\}_{\ell=1}^{\infty}$,
for $i=1, 2, \ldots, d$, are stationary, independent, and identically
distributed binary sequence with \mbox{$\Prob(\binsdmes_{i}(\ell) = 1)
  = \avesdmes_i$}. Then assuming $\prod_{i=1}^{d} \avesdmes_i +
\prod_{i=1}^{d} (1-\avesdmes_i) > 0$, the binary sequence $\binsdmarg
= \binequopt_{i=1}^{d} \binsdmes_i$ forms a time-reversible Markov
chain with the following properties:
\begin{enumerate}
\item[(a)] The transition probabilities are
\begin{align}
\label{EqnTransProbZeroOne}
\Prob\big(\binsdmarg(\ell) = 1 \,|\, \binsdmarg(\ell-1) = 0 \big)& \;
= \; \prod_{i=1}^{d}\avesdmes_{i}, \quad \text{and}\\
\label{EqnTransProbOneZero}
\Prob\big(\binsdmarg(\ell) = 0 \,|\, \binsdmarg(\ell-1) = 1 \big)& \;
= \; \prod_{i=1}^{d}(1-\avesdmes_{i}).
\end{align}

\item[(b)]  The stationary distribution is equal to
\begin{align*}
\lim_{\ell\to\infty}\Prob\big(\binsdmarg(\ell) = 1\big) \; = \;
\frac{\prod_{i=1}^{d}\avesdmes_{i}}{\prod_{i=1}^{d}\avesdmes_{i} +
  \prod_{i=1}^{d}(1-\avesdmes_{i})}.
\end{align*}

\end{enumerate}

\end{lemma}

\noindent The proof of this lemma is straight forward and is deferred
to Appendix~\ref{AppProofMarkovChain}. Now let
$\avesdmes_{\fvarfchk}^{\Time} =
\Expt[\binsdmes_{\fvarfchk}^{\Time}(\ell)] =
\Prob(\binsdmes_{\fvarfchk}^{\Time}(\ell) = 1)$ be the expected
message from the variable node $\fvar$ to the check node $\fchk$. By
construction and the fact that the variables
$\{\binsdmes_{\fvarfchk}^{\Time}(\ell)\}_{\ell=1}^{2\dimn}$ are
i.i.d., the expected value $\avesdmes_{\fvarfchk}^{\Time}$ is
independent of $\ell$. Similarly define
\mbox{$\avesdmes_{\fchkfvar}^{\Time} =
  \Expt[\binsdmes_{\fchkfvar}^{\Time}(\ell)] =
  \Prob(\binsdmes_{\fchkfvar}^{\Time}(\ell) = 1)$}, the expected
message from the check node $\fchk$ to variable node $\fvar$. Taking
expectation on both sides of the equation~\eqref{EqnSDmarg}, we obtain
\begin{align}\label{EqnExpMargOne}
\Expt\big[\sdmarg_{\fvar}^{\Time+1}\big] \; = \; \frac{1}{\dimn} \:
\sum_{\ell=\dimn+1}^{2\dimn}
\Prob\big(\binsdmarg_{\fvar}^{\Time+1}(\ell) \,=\, 1\big).
\end{align}
Therefore, in order to upper-bound the expected marginal
$\Expt[\sdmarg_{\fvar}^{\Time+1}]$, we need to calculate the
probabilities $\Prob(\binsdmarg_{\fvar}^{\Time+1}(\ell) \,=\, 1)$, for
$\ell=\dimn+1, \dimn+2, \ldots, 2\dimn$. From
Lemma~\ref{LemMarkovChain}, we know that the sequence
$\{\binsdmarg_{\fvar}^{\Time+1}(\ell)\}_{\ell=1}^{2\dimn}$ is a Markov
chain with the following transition probabilities: 
\begin{align}\label{EqnDefNodeFunBeta}
\tranzo(\avesdmes) \; \defn \;
\bpmes_{\fvar}\prod_{\fchk\in\neig(\fvar)}\avesdmes_{\fchkfvar}^{\Time},
\quad \text{and} \quad \tranoz(\avesdmes) \; \defn \;
(1-\bpmes_{\fvar})\prod_{\fchk\in\neig(\fvar)}(1-\avesdmes_{\fchkfvar}^{\Time}),
\end{align}
where $\tranzo(\cdot)$, and $\tranoz(\cdot)$ are multivariate
functions, taking values in the space $[0
  ,1]^{|\neig(\fvar)|}$. Recalling the basic Markov chain theory, we
can calculate the probability
$\Prob(\binsdmarg_{\fvar}^{\Time+1}(\ell) \,=\, 1)$ in terms of the
stationary distribution, the iteration number, and the second
eigenvalue\footnote{It is not hard to see that the second eigenvalue
  of the transition matrix of the Markov chain
  $\{\binsdmarg_{\fvar}^{\Time+1}(\ell)\}_{\ell=0}^{\infty}$ is equal
  to $(1 - \tranzo(\avesdmes) - \tranoz(\avesdmes))$.}  of the
transition matrix~\cite{Grimmett}. Doing some algebra, we obtain
\begin{align}\label{EqnIndivProb}
\nonumber \Prob\big(\binsdmarg_{\fvar}^{\Time+1}(\ell) = 1\big) \; = &
\; \Prob\big(\binsdmarg_{\fvar}^{\Time+1}(0) = 0\big) \:
\Prob\big(\binsdmarg_{\fvar}^{\Time+1}(\ell) = 1 \mid
\binsdmarg_{\fvar}^{\Time+1}(0) = 0\big)\\ \nonumber \, & + \,
\Prob\big(\binsdmarg_{\fvar}^{\Time+1}(0) = 1\big)\:
\Prob\big(\binsdmarg_{\fvar}^{\Time+1}(\ell) = 1 \mid
\binsdmarg_{\fvar}^{\Time+1}(0) = 1\big) \\ \nonumber \; = & \;
\Big[\frac{\tranoz(\avesdmes) \: \Prob\big(\binsdmarg_{\fvar}^{\Time+1}(0)
  = 1\big)}{\tranzo(\avesdmes) + \tranoz(\avesdmes)} -
\frac{\tranzo(\avesdmes) \: \Prob\big(\binsdmarg_{\fvar}^{\Time+1}(0) =
0\big)}{\tranzo(\avesdmes) + \tranoz(\avesdmes)}\Big] \: \big(1 -
  \tranzo(\avesdmes) - \tranoz(\avesdmes)\big)^{\ell}\\ \, &+ \,
  \frac{\tranzo(\avesdmes)}{\tranzo(\avesdmes) + \tranoz(\avesdmes)} .
\end{align}
Substituting equation~\eqref{EqnIndivProb} into~\eqref{EqnExpMargOne},
doing some algebra simplifying the expression, and exploiting the facts
\begin{align*}
\tranoz(\avesdmes) \: \Prob\big(\binsdmarg_{\fvar}^{\Time+1}(0) =
1\big) \, - \, \tranzo(\avesdmes) \:
\Prob\big(\binsdmarg_{\fvar}^{\Time+1}(0) = 0\big) \; \le \;
\tranzo(\avesdmes) + \tranoz(\avesdmes)
\end{align*}
and
\begin{align}\label{EqnLessThanOne}
\tranzo(\avesdmes) + \tranoz(\avesdmes) \; \le \; \bpmes_{\fvar} + (1
- \bpmes_{\fvar}) \; = \; 1
\end{align}
yields
\begin{align*}
\Expt\big[\sdmarg_{\fvar}^{\Time+1}\big] \; \le \;
\frac{\tranzo(\avesdmes)}{\tranzo(\avesdmes) + \tranoz(\avesdmes)} \,
+ \, \frac{1}{\dimn} \: \frac{1}{\tranzo(\avesdmes) +
  \tranoz(\avesdmes)} \, \big(1 - \tranzo(\avesdmes) -
\tranoz(\avesdmes)\big)^{\dimn+1}.
\end{align*}
On the other hand, denoting
\begin{align}\label{EqnDefNodeFunAlpha}
\tranzo(\bpmes)\defn
\bpmes_{\fvar}\prod_{\fchk\in\neig(\fvar)}\bpmes_{\fchkfvar}^{\Time}
\quad \text{and} \quad \tranoz(\bpmes)\defn
(1-\bpmes_{\fvar})\prod_{\fchk\in\neig(\fvar)}(1-\bpmes_{\fchkfvar}^{\Time}),
\end{align}
by definition we have 
\begin{align*}
\bpmarg_{\fvar}^{\Time+1} \; = \;
\frac{\tranzo(\bpmes)}{\tranzo(\bpmes) + \tranoz(\bpmes)}.
\end{align*}
Since the multivariate function $\tranzo(\cdot)/(\tranzo(\cdot) +
\tranoz(\cdot))$ is Lipschitz, assuming
\begin{align*}
\min\{\tranzo(\bpmes) + \tranoz(\bpmes), \tranzo(\avesdmes) +
\tranoz(\avesdmes)\} \; \ge \; \lbnd,
\end{align*} 
for some positive constant $\lbnd>0$, there exists a constant $\lips =
\lips(\lbnd)$ such that
\begin{align}\label{EqnMargError}
|\Expt\big[\sdmarg_{\fvar}^{\Time+1}\big] -
\bpmarg_{\fvar}^{\Time+1}|\; \le \; \lips & \:
\sum_{\fchk\in\neig(\fvar)}|\avesdmes_{\fchkfvar}^{\Time} -
\bpmes_{\fchkfvar}^{\Time}| \, + \, \frac{1}{\dimn} \:
\frac{(1-\lbnd)^{\dimn+1}}{\lbnd}.
\end{align}
Subsequently, in order to upper-bound the error we need to bound the
difference between expected stochastic messages and SP messages,
i.e. $|\avesdmes_{\fchkfvar}^{\Time} -
\bpmes_{\fchkfvar}^{\Time}|$. The following lemma, proved in
Appendix~\ref{AppProofMessBnd}, addresses this problem.
\begin{lemma}\label{LemMessBnd}
On a tree-structured factor graph and for sufficiently large $\dimn$,
there exists a fixed positive constant \mbox{$\lbnd < 1$} such that
\begin{align}\label{EqnLowBoundConst}
\min \big\{\ffun{\fvar}{\Time}{\bpmes} + \gfun{\fvar}{\Time}{\bpmes},
\ffun{\fvar}{\Time}{\avesdmes} + \gfun{\fvar}{\Time}{\avesdmes}\big\} \; \ge
\; \lbnd,
\end{align}
for all $\Time = 0, 1, 2, \ldots$, and $\fvar = 1, 2, \ldots,
\numvar$. Furthermore, denoting the maximum check and variable node
degrees by $\chkdeg$, and $\vardeg$, respectively, we have
\begin{align}\label{EqnMessBnd}
\max_{(\fchkfvar)} |\avesdmes_{\fchkfvar}^{\Time} -
\bpmes_{\fchkfvar}^{\Time}| \;\le\; \frac{(\chkdeg-1)
  (1-\lbnd)^{\dimn+1}}{\dimn \: \lbnd} \:
\frac{[\lips(\chkdeg-1)(\vardeg-1)]^{\diam}}
     {\lips(\chkdeg-1)(\vardeg-1) - 1},
\end{align}
for all $\Time= 0, 1, 2, \ldots$.
\end{lemma}

\noindent Now substituting inequality~\eqref{EqnMessBnd}
into~\eqref{EqnMargError}, we obtain
\begin{align*}
\max_{0\le\fvar\le\numvar} |\Expt\big[\sdmarg_{\fvar}^{\Time+1}\big] -
\bpmarg_{\fvar}^{\Time+1}|\; \le \; \frac{(1-\lbnd)^{\dimn+1}}{\dimn
  \: \lbnd} \, \left\{\lips (\chkdeg-1)\vardeg\:\frac{[\lips
    (\chkdeg-1)(\vardeg-1)]^{\diam}}{\lips (\chkdeg-1)(\vardeg-1)-1}
\,+\, 1\right\},
\end{align*}
for all $\Time = 0, 1, 2, \ldots$. Therefore, setting
\begin{align}\label{EqnWhatIsK}
\dimn \;=\; \max\left\{\frac{\log{\accu} - \diam \: \log(\lips
  (\chkdeg-1)(\vardeg-1))}{\log(1 - \lbnd)} \, , \, \frac{3}{\lbnd}
\right\},
\end{align}
we obtain
\begin{align*}
\max_{0\le\fvar\le\numvar} |\Expt\big[\sdmarg_{\fvar}^{\Time}\big] -
\bpmarg_{\fvar}^{\Time}|\; \le \; \accu,
\end{align*}
for all $\Time=0, 1, 2, \ldots$.


\subsection{Proof of Part $(b)$ of Theorem~\ref{ThmTree}}
\label{SubSecPartB}

To stramline the exposition, let $\binsdmarg(\ell) \defn
\binsdmarg_{\fvar}^{\Time+1}(\ell)$, for fixed $\fvar$, and
$\Time$. As previously stated, the sequence
$\{\binsdmarg(\ell)\}_{\ell=1}^{\infty}$ is a Markov chain with
initial state $p_0 \defn \Prob(\binsdmarg(0) = 0)$, $p_1 \defn
\Prob(\binsdmarg(0) = 1)$, and transition probabilities $\fsim \defn
\ffun{\fvar}{\Time}{\avesdmes}$, and $\gsim \defn
\gfun{\fvar}{\Time}{\avesdmes}$; more specifically we have
\begin{align*}
\fsim \; = \;\Prob\big(\binsdmarg(\ell) = 1 \,|\, \binsdmarg(\ell-1) =
0 \big), \; \quad \text{and} \quad \gsim \;=\;
\Prob\big(\binsdmarg(\ell) = 0 \,|\, \binsdmarg(\ell-1) = 1 \big).
\end{align*}
Since $\Expt[(\binsdmarg(\ell)-\Expt[\binsdmarg(\ell)])^2] \le 1$, in
order to upper-bound the variance
\begin{align*}
\var\big(\sdmarg_{\fvar}^{\Time+1}\big) \; = \; \frac{1}{\dimn^2}
\sum_{\ell=\dimn+1}^{2\dimn} \Expt\big[\big(\binsdmarg(\ell) -
  \Expt[\binsdmarg(\ell)]\big)^2\big] \, + \, \frac{2}{\dimn^2}
\sum_{\ell' < \ell} \Expt\big[\big(\binsdmarg(\ell) -
  \Expt[\binsdmarg(\ell)]\big) \big(\binsdmarg(\ell') -
  \Expt[\binsdmarg(\ell')]\big)\big],
\end{align*}
we only need to upper-bound the cross-product terms. Doing so, for
$\ell > \ell'$, we have
\begin{align*}
\Expt\big[\big(\binsdmarg(\ell) - \Expt[\binsdmarg(\ell)]\big)
  \big(\binsdmarg(\ell') - \Expt[\binsdmarg(\ell')]\big)\big] \; = &\;
\Expt\big[\binsdmarg(\ell)\:\binsdmarg(\ell')\big] \, - \,
\Expt\big[\binsdmarg(\ell)\big] \: \Expt\big[\binsdmarg(\ell')\big]
\\ \; = &\; \Prob\big(\binsdmarg(\ell') = 1\big) \:
\big[\Prob\big(\binsdmarg(\ell) = 1 | \binsdmarg(\ell') = 1 \big)
  \,-\, \Prob\big(\binsdmarg(\ell) = 1\big)\big]\\ \; \le& \;
\Prob\big(\binsdmarg(\ell) = 1 | \binsdmarg(\ell') = 1 \big) \,-\,
\Prob\big(\binsdmarg(\ell) = 1\big).
\end{align*}
Now, exploiting the Markov property and equation~\eqref{EqnIndivProb},
we can further simplify the aforementioned inequality
\begin{align*}
\Expt\big[\big(\binsdmarg(\ell) - \Expt[\binsdmarg(\ell)]\big) &
  \big(\binsdmarg(\ell') - \Expt[\binsdmarg(\ell')]\big)\big]\\ \; &
\le \; \frac{\gsim}{\fsim+\gsim} \: (1 - \fsim -\gsim)^{(\ell-\ell')}
\, + \, \frac{\fsim\:p_0}{\fsim + \gsim} \: (1 - \fsim
-\gsim)^{\ell}\, - \, \frac{\gsim\:p_1}{\fsim + \gsim} \: (1 - \fsim
-\gsim)^{\ell}\\ \; & \stackrel{\text{(i)}}{\le} \; (1 - \fsim
-\gsim)^{(\ell-\ell')},
\end{align*}
where inequality (i) follows from~\eqref{EqnLessThanOne} and the fact
that $\gsim/(\fsim+\gsim) + p_0\fsim/(\fsim+\gsim) \le 1$. According
to Lemma~\ref{LemMessBnd}, for sufficiently large $\dimn$, we have
$\fsim + \gsim \ge \lbnd$. Therefore, putting the pieces together
doing some algebra, we obtain
\begin{align*}
\var\big(\sdmarg_{\fvar}^{\Time+1}\big) \; \le& \; \frac{1}{\dimn}\, +
\, \frac{2}{\dimn^2} \sum_{\dimn+1 \le \ell' < \ell\le 2\dimn} (1 -
\lbnd)^{(\ell-\ell')} \\ \; =& \; \frac{1}{\dimn}\, + \,
\frac{2}{\dimn^2} \sum_{i=1}^{\dimn-1} (\dimn-i)(1 -
\lbnd)^{i}\\ \;\le&\; \frac{1}{\dimn}\, + \, \frac{2}{\dimn}
\sum_{i=1}^{\dimn-1} (1 - \lbnd)^{i} \; \le \; \frac{1 +
  2/\lbnd}{\dimn},
\end{align*}
for all $\fvar = 1, 2, \ldots, \numvar$, and $\Time = 0, 1, 2,
\ldots$.


\section{Experimental Results} 
\label{SecSimulations}

\begin{figure}
\begin{center}
\begin{tabular}{cc}
  \widgraph{.47\textwidth}{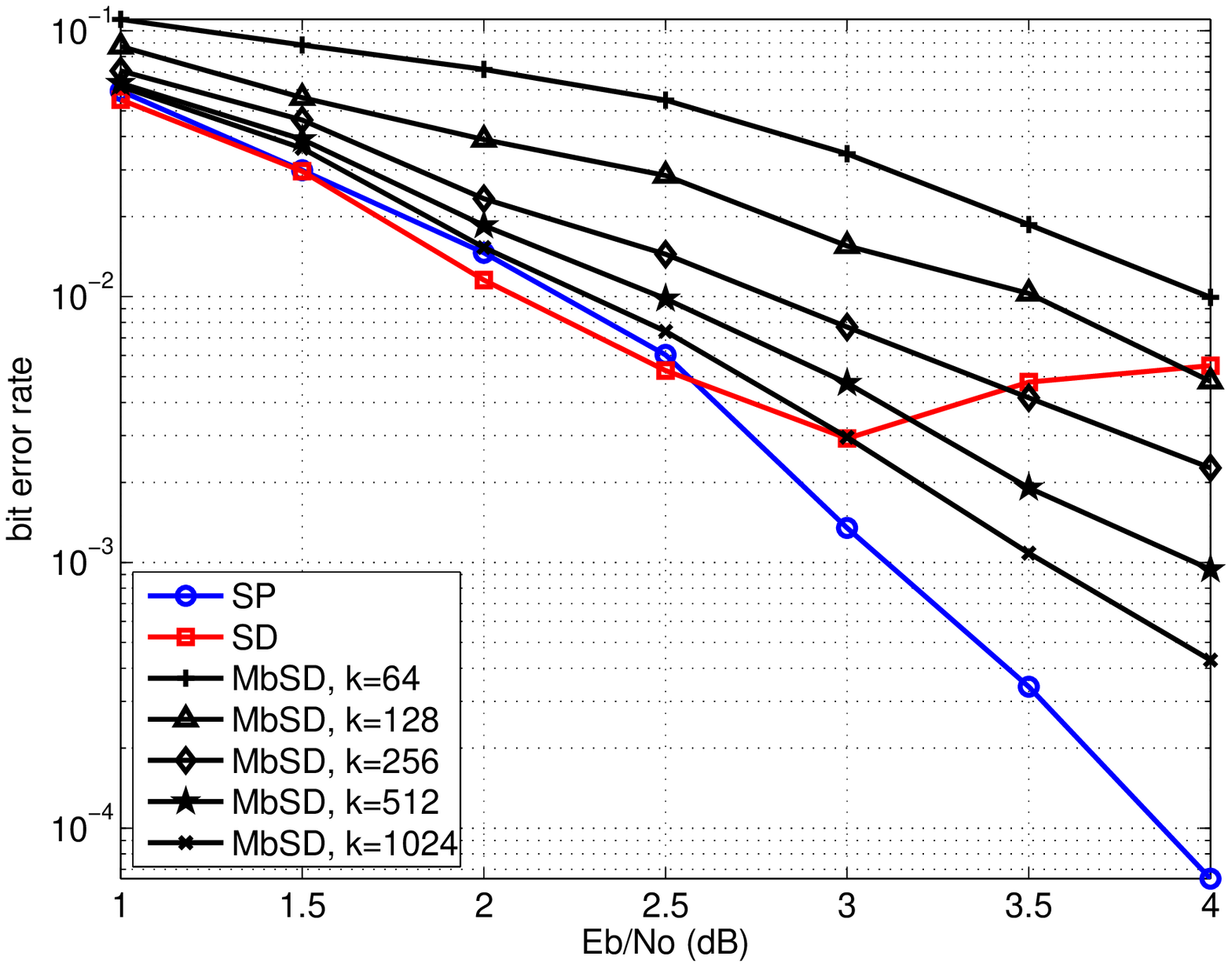} &
  \widgraph{.47\textwidth}{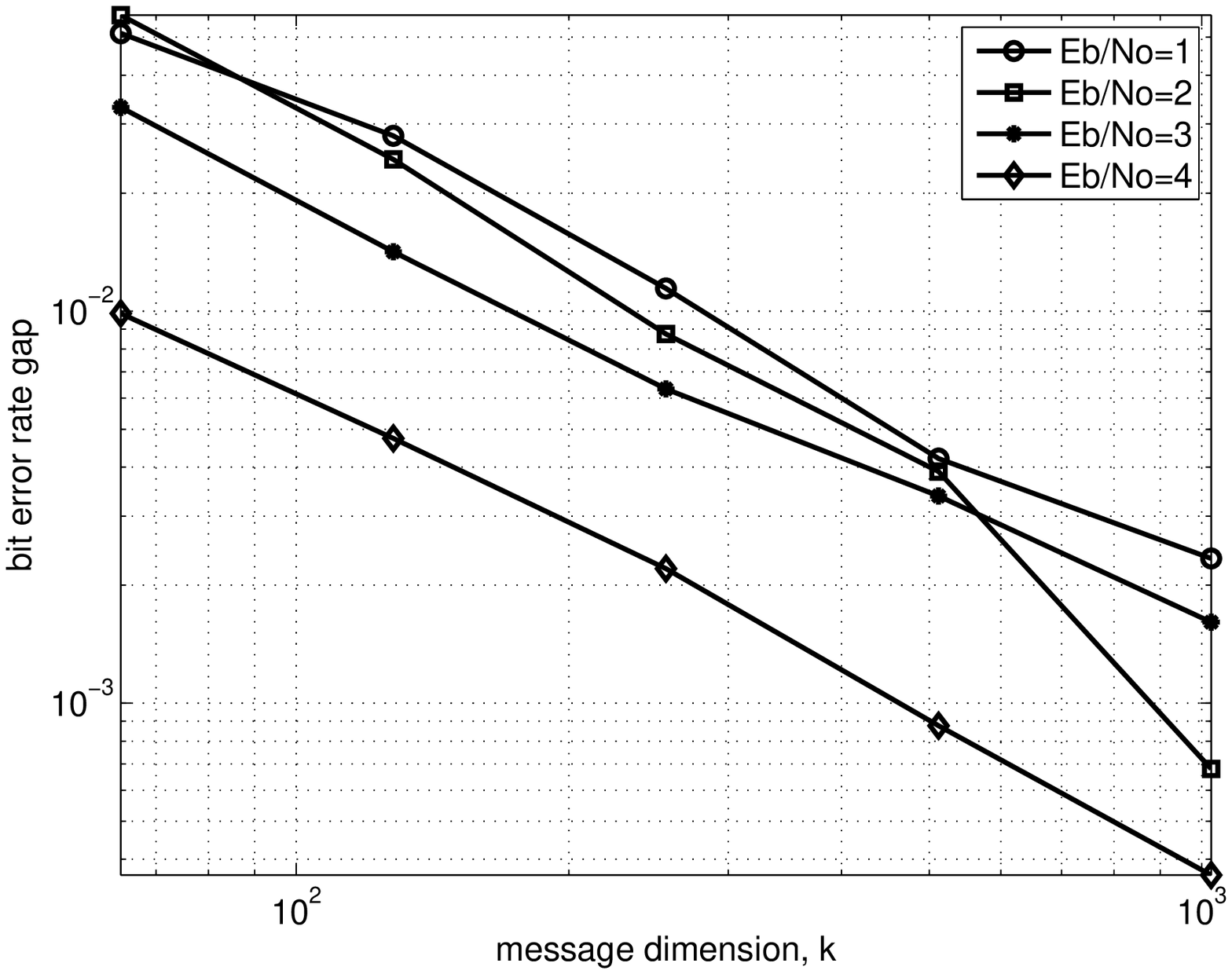}\\ (a) & (b)
\end{tabular}
\end{center}
\caption{Performance of the \MySD algorithm on a (3,6)-LDPC code with
  $\numvar=200$ variable nodes and $\numchk=100$ check nodes. (a) Bit
  error rate versus the energy per bit to noise power spectral density
  (Eb/No) for different decoders, namely, the SP, the SD (stochastic
  decoding without noise dependent scaling~\cite{TehraniEtal06}), and
  the \MySD for different message dimensions $\dimn\in\{64, 128, 256,
  512, 1024\}$. As predicted by the theory, the performance of the
  \MySD converges to that of SP. Moreover, \MySD does not suffer from
  error floor, in contrast to the SD algorithm. (b) Bit error rate
  gap, i.e., the difference between the SP and the \MySD bit error
  rates, versus the message dimension. The rate of convergence is
  upper bounded by $\order(1/\dimn)$, manifested as linear curves in
  the log-log domain plot.}\label{FigBER_200100}
\end{figure}

To confirm our theoretical predictions, we test the \MySD algorithm on
a simple LDPC code. In our experiments, we set the block-length
(variable nodes) and number of parity checks (check nodes) to be
$\numvar = 200$ and $\numchk = 100$, respectively. Using Gallager's
construction~\cite{RicUrb}, we first generate a regular (3, 6)-LDPC
code, that is, all variable nodes have degree three, whereas all check
nodes have degree six. Then, considering a binary pulse-amplitude
modulation (BPAM) system over an AWGN channel, we run MbSD, for $\Time
= 60$ iterations, on several simulated signals in order to compute the
bit error rate for different values of normalized signal to noise
ratio.\footnote{A BPAM system with transmit power one over an AWGN
  channel with noise variance $\sigma^2$ has the energy per bit to
  noise power spectral density (Eb/No) of $1/(2R\sigma^2)$, where $R$
  is the code rate.} The test is carried out for a number of message
dimensions $\dimn$, and results are compared with the SP algorithm and
the stochastic decoding (SD) proposed by Tehrani et
al.~\cite{TehraniEtal06} (see Figure~\ref{FigBER_200100}
(a)).\footnote{In simulating the SD algorithm, we used 30000 `decoding
  cycles' and edge memory of length 25 without noise dependent
  scaling.} As predicted by our theorems, the performance of the \MySD
converges to that of SP as $\dimn$ grows. Therefore, in contrast to
the SD algorithm, \MySD is an asymptotically consistent estimate of
the SP and does not suffer from error floor. The rate of convergence,
on the other hand, can be further explored in
Figure~\ref{FigBER_200100} (b), wherein the bit error rate gap (i.e.,
the difference between SP and \MySD bit error rates) versus the
message dimension is illustrated. As can be observed, the error curves
in the log-log domain plot are roughly linear with slope one. This
observation is consistent with equation~\eqref{EqnConvRate},
suggesting the upper-bound of $\order(1/\dimn)$ for the rate of
convergence.

\begin{figure}
\begin{center}
  \widgraph{.63\textwidth}{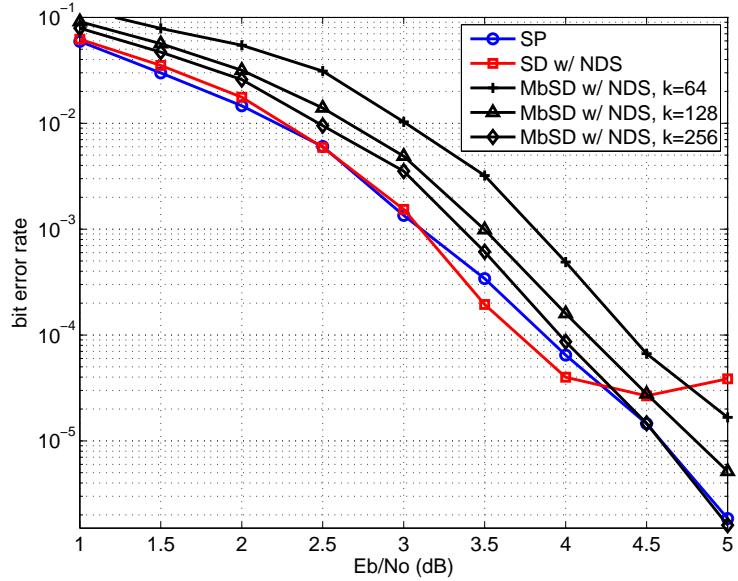}
\end{center}
\caption{Effect of the noise dependent scaling (NDS) on the
  performance of the \MySD algorithm. The panel contains several plots
  illustrating the bit error rate versus the energy per bit to noise
  power spectral density (Eb/No) for different decoders, namely, the
  SP, the SD (stochastic decoding with noise dependent
  scaling~\cite{TehraniEtal06}), and the \MySD for different message
  dimensions $\dimn\in\{64, 128, 256\}$. As expected, NDS has
  significantly improved the performance of the MbSD. Simulations were
  conducted on a (3,6)-LDPC code with $\numvar=200$ variable nodes and
  $\numchk=100$ check nodes.}\label{FigBER_NDS_200100}
\end{figure}

To improve upon the seemingly slow rate of convergence, we make use of
the notion of noise dependent scaling (NDS). Sensitivity to random
switching activities, referred to as `latching', has been observed to
be a major challenge in stochastic
decoders~\cite{WinsteadEtal05,TehraniEtal06}. To circumvent this
issue, the notion of NDS, in which the received log-likelihoods are
down-scaled by a factor proportional to the SNR, was proposed and
shown to be extremely effective~\cite{TehraniEtal06}. The \MySD
algorithm suffers from the latching problem too, especially for high
SNR values. Intuitively, sequences generated by Markov chains (recall
Lemma~\ref{LemMarkovChain}) are likely to be the all-one or the
all-zero sequences when the SNR is sufficiently high. As a
consequence, in such cases, the positive constant $\lbnd$, defined
in~\eqref{EqnLowBoundConst}, is more likely to be close to zero. The
rate of convergence of the expectation, specified in
equation~\eqref{EqnWhatIsK}, is inversely proportional to
$\log(1-\lbnd)$, therefore, the smaller the $\lbnd$, the slower the
rate of convergence. Resolving this issue requires increasing the
switching activities of Markov chain, which is accomplished by the
NDS. Figure~\ref{FigBER_NDS_200100}, illustrates the bit error rate
versus Eb/No for the SP, the SD using NDS, and the \MySD using
NDS.\footnote{In our simulations we set the NDS scaling parameter to
  be $\sigma^2$, the optimum choice as suggested in the
  paper~\cite{TehraniEtal06}.} As is evident, the rate of convergence
of the \MySD algorithm, and thus its performance, is significantly
improved. Moreover, having the same number of decoding cycles, \MySD
outperforms the SD for high SNRs.

\section{Conclusion}
\label{SecConclusion}

In this paper, we studied the theoretical aspect of stochastic
decoders, a widely studied solution for fully-parallel implementation
of LDPC decoding on chips. Generally speaking, encoding messages by
binary sequences, stochastic decoders simplify check and node message
updates by modulo-two summation and the equality operator,
respectively. As it turns out, the check node operation is
statistically consistent on average, whereas, the variable node
equality operation generates a Markov chain with the desired quantity
as its stationary distribution. Therefore, for a finite message
dimension $\dimn$, the stochastic message updates introduce error
terms which become propagated in the factor graph. Controlling these
errors is the main challenge in the theoretical analysis of stochastic
decoders. To formalize these notions, we introduced a novel stochastic
algorithm, referred to as the Markov based stochastic decoding, and
provided concrete theoretical guarantees on its performance. More
precisely, we showed that expected marginals produced by the \MySD
become arbitrarily close to marginals generated by the SP algorithm on
tree-structured as well as general factor graphs. The rate of
convergence is governed by the message dimension, the graph structure,
and the Lipschitz constant, formally specified in
equation~\eqref{EqnWhatIsK}. Moreover, we proved that the variance of
\MySD marginals are upper-bounded by $\order(1/\dimn)$. These
theoretical predictions were also supported by experimental
results. We showed that, maintaining the same order of decoding
cycles, our algorithm does not suffer from error floor; therefore, it
achieves better bit error rate performance compared to other competing
methods.

\subsection*{Acknowledgements}

Authors would like to thank Aman Bahtia for providing the C++ code,
simulating sum-product and stochastic decoding algorithms.


\appendix


\section{Proof of Lemma~\ref{LemMarkovChain}}
\label{AppProofMarkovChain}

By definition we have
\begin{align*}
\binsdmarg(\ell) \; = \; \left\{\begin{array}{lc} 
1 & \text{if $\binsdmes_{\fvar}(\ell) = 1$, for all $i = 1, 2, \ldots, d$}\\ 
0 & \text{if $\binsdmes_{\fvar}(\ell) = 1$, for all $i = 1, 2, \ldots, d$}\\
\binsdmarg(\ell-1) & \text{otherwise}
\end{array}\right..
\end{align*}
Therefore, given $\binsdmarg(\ell-1)=0$ and regardless of the sequence
$\{\binsdmarg(\ell-2), \ldots, \binsdmarg(0)\}$, the event
$\{\binsdmarg(\ell) = 1\}$ is equivalent to $\{\binsdmes_{\fvar}(\ell)
= 1, \forall \, \fvar=1, \ldots, d\}$. Therefore, we have
\begin{align*}
\Prob\big(\binsdmarg(\ell)=1 \, | \, \binsdmarg(\ell-1) = 0,
\binsdmarg(\ell-2), \ldots\big) \; & = \;
\Prob\big(\binsdmes_{\fvar}(\ell) = 1, \, \forall \fvar = 1, \ldots,
d,\big)\\ \; & \stackrel{(\text{i})}{=}
\;\prod_{\fchk=1}^{d}\Prob\big(\binsdmes_{\fvar}(\ell) = 1\big) \;= \;
\prod_{\fvar=1}^{d}\avesdmes_{\fvar},
\end{align*}
where equality (i) follows from the i.i.d. nature of the sequence. The
exact same argument yields the
equation~\eqref{EqnTransProbOneZero}. Finally, the stationary
distribution can be obtained from the detailed balance
condition~\cite{Grimmett}
\begin{align*}
\lim_{\ell\to\infty}\Prob\big(\binsdmarg_{\fvar}(\ell) = 1\big) \:
\prod_{\fvar=1}^{d}(1-\avesdmes_{\fvar}) \; = \; \lim_{\ell\to\infty}
\Prob\big(\binsdmarg_{\fvar}(\ell) = 0\big) \:
\prod_{\fvar=1}^{d}\avesdmes_{\fvar}.
\end{align*}


\section{Proof of Lemma~\ref{LemMessBnd}}
\label{AppProofMessBnd}
Recall binary messages $\binsdmes_{\fchkfvar}^{\Time}$, and
$\binsdmes_{\fvarfchk}^{\Time}$ from steps 2(a) and 2(b) of the main
algorithm. Also recall the definition of expected messages
$\avesdmes_{\fchkfvar}^{\Time} =
\Prob(\binsdmes_{\fchkfvar}^{\Time}(\ell) = 1)$, and
$\avesdmes_{\fvarfchk}^{\Time} =
\Prob(\binsdmes_{\fvarfchk}^{\Time}(\ell) = 1)$. From Lemma 1 of the
paper~\cite{Gallager62}, we know that
\begin{align}\label{EqnSDexpMesChk2Var}
\avesdmes_{\fchkfvar}^{\Time} \; = \; \frac{1}{2} \, - \, \frac{1}{2} \:
\prod_{\compneigchk}\big(1 - 2\avesdmes_{\svarfchk}^{\Time}\big).
\end{align}
On the other hand, by construction we have 
\begin{align*}
\avesdmes_{\fvarfchk}^{\Time+1} \; = \; \frac{1}{\dimn} \:
\sum_{\ell=\dimn+1}^{2\dimn} \Prob\big(\auxsdmes_{\fvarfchk}^{\Time+1}(\ell) = 1\big),
\end{align*}
where $\auxsdmes_{\fvarfchk}^{\Time+1} \;\; =
\binequopt_{\compneigvar} \binsdmes_{\schkfvar}^{\Time} \; \binequopt
\; \binsdmes_{\fvar}^{\Time+1}$. Since, according to
Lemma~\ref{LemMarkovChain}, the sequence
$\{\auxsdmes_{\fvarfchk}^{\Time+1}(\ell)\}$ forms a Markov chain with
transition probabilities
\begin{align*} 
\tranzoEdge(\avesdmes) \; \defn \;
\bpmes_{\fvar}\prod_{\compneigvar}\avesdmes_{\schkfvar}^{\Time}, \quad
\text{and} \quad \tranozEdge(\avesdmes) \; \defn \;
(1-\bpmes_{\fvar})\prod_{\compneigvar}(1-\avesdmes_{\schkfvar}^{\Time}), 
\end{align*}
basic Markov chain theory yields
\begin{align*}
\Prob&\big(\auxsdmes_{\fvarfchk}^{\Time+1}(\ell) = 1\big) \; = \;
\frac{\tranzoEdge(\avesdmes)}{\tranzoEdge(\avesdmes) +
  \tranozEdge(\avesdmes)} \\ \, & + \,
\Big[\frac{\tranozEdge(\avesdmes)
    \Prob\big(\auxsdmes_{\fvarfchk}^{\Time+1}(0)=1\big)}{\tranzoEdge(\avesdmes)
    + \tranozEdge(\avesdmes)} - \frac{\tranzoEdge(\avesdmes)
    \Prob\big(\auxsdmes_{\fvarfchk}^{\Time+1}(0)=0\big)}{\tranzoEdge(\avesdmes)
    + \tranozEdge(\avesdmes)} \Big]\: \big(1 - \tranzoEdge(\avesdmes)
- \tranozEdge(\avesdmes)\big)^{\ell}.
\end{align*}
Therefore, doing some algebra, noticing the facts that
\begin{align*}
\frac{\tranozEdge(\avesdmes)
  \Prob\big(\auxsdmes_{\fvarfchk}^{\Time+1}(0)=1\big)}{\tranzoEdge(\avesdmes)
  + \tranozEdge(\avesdmes)} - \frac{\tranzoEdge(\avesdmes)
  \Prob\big(\auxsdmes_{\fvarfchk}^{\Time+1}(0)=0\big)}{\tranzoEdge(\avesdmes)
  + \tranozEdge(\avesdmes)} \; \le \; 1
\end{align*}
and 
\begin{align*}
\tranzoEdge(\avesdmes) + \tranozEdge(\avesdmes) \; \le \;
\bpmes_{\fvar} + (1 - \bpmes_{\fvar}) \; = \; 1,
\end{align*}
we have
\begin{align}\label{EqnSDexpMesVar2Chk}
\avesdmes_{\fvarfchk}^{\Time+1} \; \le \;
\frac{\tranzoEdge(\avesdmes)}{\tranzoEdge(\avesdmes) +
  \tranozEdge(\avesdmes)} \, + \,
\frac{1}{\dimn\:(\tranzoEdge(\avesdmes) + \tranozEdge(\avesdmes))} \:
(1 - \tranzoEdge(\avesdmes) - \tranozEdge(\avesdmes))^{\dimn+1}.
\end{align}
Equations~\eqref{EqnSDexpMesChk2Var}, and~\eqref{EqnSDexpMesVar2Chk}
characterize the stochastic decoding message updates. Similarly, we
have the SP update equations for all directed edges
\begin{align}\label{EqnBPmesUpdate}
\bpmes_{\fchkfvar}^{\Time} \; = \; \frac{1}{2} \, - \, \frac{1}{2} \:
\prod_{\compneigchk}\big(1 - 2\bpmes_{\svarfchk}^{\Time}\big), \quad
\text{and} \quad \bpmes_{\fvarfchk}^{\Time+1} \; = \;
\frac{\tranzoEdge(\bpmes)}{\tranzoEdge(\bpmes) + \tranozEdge(\bpmes)},
\end{align}
where we have denoted 
\begin{align*}
\tranzoEdge(\bpmes)\; \defn \;
\bpmes_{\fvar}\prod_{\compneigvar}\bpmes_{\schkfvar}^{\Time}, \quad
\text{and} \quad \tranozEdge(\bpmes) \; \defn \;
(1-\bpmes_{\fvar})\prod_{\compneigvar}(1-\bpmes_{\schkfvar}^{\Time}).
\end{align*}
Since $ 0 \le \bpmes_{\svarfchk}^{\Time}\le 1$, and $0 \le
\avesdmes_{\svarfchk}^{\Time}\le 1$, for all $\Time$, and
$(\svarfchk)$, we have\footnote{The inequality follows from the
  mean-value theorem and the fact that for the function \mbox{$h(x_1,
    \ldots, x_d) = \prod_{i=1}^{d}(1-2x_i) / 2$}, we have $|\partial h
  / \partial x_i| \le 1$, if $0 < x_i \le 1$ for all $i=1,\ldots,d$. }
\begin{align}\label{EqnChk2VarBnd}
|\avesdmes_{\fchkfvar}^{\Time+1} - \bpmes_{\fchkfvar}^{\Time+1}|  \;\;\le
\sum_{\compneigchk} |\avesdmes_{\svarfchk}^{\Time+1} - \bpmes_{\svarfchk}^{\Time+1}|.
\end{align}
We now turn to upper-bounding the term
\begin{align*}
|\avesdmes_{\svarfchk}^{\Time+1} \, - \, \bpmes_{\svarfchk}^{\Time+1}|
\; \le & \; \Big|
\frac{\ffun{\svarfchk}{\Time}{\avesdmes}}{\ffun{\svarfchk}{\Time}{\avesdmes}
  + \gfun{\svarfchk}{\Time}{\avesdmes}} \, - \,
\frac{\ffun{\svarfchk}{\Time}{\bpmes}}{\ffun{\svarfchk}{\Time}{\bpmes}
  + \gfun{\svarfchk}{\Time}{\bpmes}} \Big| \\ & +
\frac{1}{\dimn\:(\ffun{\svarfchk}{\Time}{\avesdmes} +
  \gfun{\svarfchk}{\Time}{\avesdmes})} \: (1 -
\ffun{\svarfchk}{\Time}{\avesdmes} -
\gfun{\svarfchk}{\Time}{\avesdmes})^{\dimn+1}.
\end{align*}

\noindent Since $0 < \bpmes_{\svarfchk}^{0} = \bpmes_{\svar} <
1$. Then by inspection of the SP updates, it is easy to see that $0<
\bpmes_{\svarfchk}^{\Time} <1$, and $0 <\bpmes_{\fchksvar}^{\Time}
<1$, for all $\Time\ge0$, and all directed edges $(\fchksvar)$, and
$(\svarfchk)$. Therefore, recalling the
definition~\eqref{EqnDefNodeFunAlpha}, and the fact that
$\bpmes_{\fchkfvar}^{\Time} = \bpmes_{\fchkfvar}^{\ast}$, for $\Time$
larger than the graph diameter, there exists a positive constant
$\lbnd < 1$ such that
\begin{align*}
\ffun{\svar}{\Time}{\bpmes} \, + \, \gfun{\svar}{\Time}{\bpmes} \; \ge \;
2\:\lbnd, \quad \text{for all} \quad \svar = 1, 2, \ldots, \numvar, \;
\text{and} \;\; \Time = 0, 1, 2, \ldots.
\end{align*}
Now we show that for sufficiently large $\dimn$, we have 
\begin{align}\label{EqnClaim}
\ffun{\svar}{\Time}{\avesdmes} \, + \, \gfun{\svar}{\Time}{\avesdmes}
\; \ge \; \lbnd, \quad \text{for all} \quad \svar = 1, 2, \ldots,
\numvar, \; \text{and} \;\; \Time = 0, 1, 2, \ldots.
\end{align}
Suppose for a fixed $\Time$, we have
\begin{align}\label{EqnCondition}
\sum_{\fchk\in\neig(\svar)} |\avesdmes_{\fchksvar}^{\tau} -
\bpmes_{\fchksvar}^{\tau}| \; \le \; \lbnd, \quad \text{for all} \quad
\svar = 1, 2, \ldots, \numvar, \; \text{and} \;\; \tau = 0, 1,
\ldots, \Time.
\end{align}
Since $\avesdmes_{\svarfchk}^{0} = \bpmes_{\svarfchk}^{0} =
\bpmes_{\svar}$, the left hand side of the above inequality is
initially equal to zero; thus the condition~\eqref{EqnCondition} is
satisfied for $\Time=0$. Now making use of the mean-value theorem, we
obtain\footnote{Let $h(x_1, \ldots, x_d) = \alpha \prod_{i=1}^{d}x_i +
  (1-\alpha)\prod_{i=1}^{d}(1-x_i)$ be an arbitrary function for $0\le
  x_i\le1$, $i=1,\ldots,d$. Then we have $|\partial h/ \partial x_i|
  \le \alpha \prod_{j\neq i}x_j + (1 - \alpha)\prod_{j\neq i}(1 - x_j)
  \le 1$. }
\begin{align*}
|\ffun{\svar}{\tau}{\avesdmes} + \gfun{\svar}{\tau}{\avesdmes} -
\ffun{\svar}{\tau}{\bpmes} - \gfun{\svar}{\tau}{\bpmes}| \; \le
\; \sum_{\fchk\in\neig(\svar)} |\avesdmes_{\fchksvar}^{\tau} -
\bpmes_{\fchksvar}^{\tau}|.
\end{align*}
Putting the pieces together, assuming~\eqref{EqnCondition}, yields
\begin{align*}
\ffun{\svar}{\tau}{\avesdmes} \, + \, \gfun{\svar}{\tau}{\avesdmes} \; \ge
\; \lbnd\quad \text{for all} \quad \svar = 1, 2, \ldots, \numvar, \;
\text{and} \;\; \tau = 0, 1, \ldots, \Time.
\end{align*}
Since $0 \le \avesdmes_{\fchksvar}^{\Time},
\bpmes_{\fchksvar}^{\Time} \le 1$, we have 
\begin{align*}
\ffun{\svarfchk}{\tau}{\avesdmes} \, + \,
\gfun{\svarfchk}{\tau}{\avesdmes} \; & \ge \;
\ffun{\svar}{\tau}{\avesdmes} \, + \, \gfun{\svar}{\tau}{\avesdmes}
\; \ge \; \lbnd, \quad \text{and}\\
\ffun{\svarfchk}{\tau}{\bpmes} \, + \,
\gfun{\svarfchk}{\tau}{\bpmes} \; & \ge \;
\ffun{\svar}{\tau}{\bpmes} \, + \, \gfun{\svar}{\tau}{\bpmes}
\; \ge \; 2\:\lbnd;
\end{align*}
therefore, there exist a constant $\lips = \lips(\lbnd)$ such that
\begin{align}\label{EqnVar2ChkBnd}
|\avesdmes_{\svarfchk}^{\tau+1} - \bpmes_{\svarfchk}^{\tau+1}| \;
\le \; \lips \sum_{\schk \in\neig(\svar)\setminus\{\fchk\} }
|\avesdmes_{\schksvar}^{\tau} - \bpmes_{\schksvar}^{\tau}| \, + \,
\frac{1}{\dimn} \frac{(1-\lbnd)^{\dimn+1}}{\lbnd},
\end{align}
for all $\tau = 0, 1, \ldots, \Time$. Substituting the
inequality~\eqref{EqnVar2ChkBnd} into~\eqref{EqnChk2VarBnd}, we obtain
\begin{align}\label{EqnRecursion}
|\avesdmes_{\fchkfvar}^{\tau+1} - \bpmes_{\fchkfvar}^{\tau+1}| \;\;
\le \; \lips \sum_{\compneigchk} \sum_{\schk
  \in\neig(\svar)\setminus\{\fchk\} } |\avesdmes_{\schksvar}^{\tau} -
\bpmes_{\schksvar}^{\tau}| \, + \, \frac{(\chkdeg-1)
  (1-\lbnd)^{\dimn+1}}{\dimn \: \lbnd}, 
\end{align}
where we have denoted $\chkdeg \defn \max_{0\le \fchk\le\numchk}
|\neig(\fchk)|$. Let $\error^{\Time} \defn
\{|\avesdmes_{\fchkfvar}^{\Time} -
\bpmes_{\fchkfvar}^{\Time}|\}_{(\fchkfvar)}$ be the $\numedge
\defn \sum_{\fchk=1}^{\numchk}|\neig(\fchk)|$ dimensional error
vector. Now define a matrix $\mat\in\real^{\numedge\times\numedge}$
with entries indexed by pairs of directed edges $(\fchkfvar)$; in
particular, we have
\begin{align}\label{EqnDefnMat}
\mat(\fchkfvar, \schksvar) \; \defn \; \left\{\begin{array}{cc}\lips &
\text{if}\;\compneigchk\;
\text{and}\;\schk\in\neig(\svar)\setminus\{\fchk\}\\ 0 & \text{o.w.}
\end{array}\right..
\end{align}
Then by stacking the scalar inequalities~\eqref{EqnRecursion}, we
obtain the vector inequality
\begin{align}\label{EqnVectorRecursion}
\error^{\tau+1} \; \vecineq \; \mat \: \error^{\tau} \, + \,
\frac{(\chkdeg-1) (1-\lbnd)^{\dimn+1}}{\dimn \: \lbnd} \: \onevec,
\end{align}
for all $\tau = 0, 1, \ldots, \Time$. Here, $\vecineq$ denotes the
vector inequality, i.e., for $\numedge$-dimensional vectors $x$, and
$y$ we say $x \vecineq y$ if and only if $x(i) \le y(i)$, for all
$i=1, 2, \ldots, \numedge$. Unwrapping the
recursion~\eqref{EqnVectorRecursion}, noticing that $\error^{0} =
\zerovec$, we obtain 
\begin{align}\label{EqnUnwrapped}
\error^{\Time+1} \; \vecineq \; \frac{(\chkdeg-1)
  (1-\lbnd)^{\dimn+1}}{\dimn \: \lbnd} \, \big(I \, + \, \mat \, + \,
\ldots \, + \, \mat^{\Time}\big) \: \onevec.
\end{align}
The right hand side sequence of the previous inequality have,
seemingly, a growing number of terms as $\Time \to \infty$. However,
according to the following lemma, proved in
Appendix~\ref{AppProofNilpotent}, the graph-respecting matrix $\mat$
is nilpotent, i.e., there exists a positive integer $\ell$ such that
$\mat^{\ell} = 0$. (A similar statement regarding nilpotence of the
tree-structured Markov random field have been shown in Lemma
1~\cite{NooWai13a}.) Recall the definition of the factor graph
diameter $\diam$, the largest path between any pair of variable nodes.
\begin{lemma}\label{LemNilpotent}
The graph-respecting matrix $\mat$, defined in~\eqref{EqnDefnMat}, is
nilpotent with degree at most the diameter of the factor graph
$\diam$, that is $\mat^{\diam} = 0$.
\end{lemma}

\noindent Exploiting the result of Lemma~\ref{LemNilpotent}, we can
further simplify the vector inequality~\eqref{EqnUnwrapped}
\begin{align}\label{EqnUnwrappedSimple}
\error^{\Time} \; \vecineq \; \frac{(\chkdeg-1)
  (1-\lbnd)^{\dimn+1}}{\dimn \: \lbnd} \, \big(I \, + \, \mat \, + \,
\ldots \, + \, \mat^{\diam-1}\big) \: \onevec,
\end{align}
for all $\Time = 0, 1, 2, \ldots$. Now we take $\vnorm{\cdot}{\infty}$
on the both sides of the
inequality~\eqref{EqnUnwrappedSimple}. Recalling the definition of the
matrix norm infinity\footnote{Norm infinity of a matrix is the maximum
  absolute row sum of the matrix~\cite{Horn85}},
$\matsnorm{\cdot}{\infty}$, triangle inequality, and using the fact
that
\mbox{$\matsnorm{\mat^{\ell}}{\infty}\le\matsnorm{\mat}{\infty}^{\ell}$}
~\cite{Horn85}, simple algebra yields
\begin{align*}
\max_{(\fchkfvar)} |\avesdmes_{\fchkfvar}^{\Time+1} -
\bpmes_{\fchkfvar}^{\Time+1}| \; & \le \; \frac{(\chkdeg-1)
  (1-\lbnd)^{\dimn+1}}{\dimn \: \lbnd} \; \sum_{\ell=0}^{\diam-1}
\matsnorm{\mat}{\infty}^{\ell} \\ \; & \le \;
\frac{(\chkdeg-1)(1-\lbnd)^{\dimn+1}}{\dimn \: \lbnd} \;
\sum_{\ell=0}^{\diam-1} [\lips(\chkdeg-1)(\vardeg-1)]^{\ell}\\ \;& \le\;
\frac{(\chkdeg-1) (1-\lbnd)^{\dimn+1}}{\dimn \: \lbnd} \:
\frac{[\lips(\chkdeg-1)(\vardeg-1)]^{\diam}}
     {\lips(\chkdeg-1)(\vardeg-1) - 1}, 
\end{align*}
where we have denoted $\vardeg = \max_{0\le \fvar\le\numvar}
|\neig(\fvar)|$. For sufficiently large $\dimn$, specifically when 
\begin{align*}
\frac{(\chkdeg-1) (1-\lbnd)^{\dimn+1}}{\dimn \: \lbnd} \:
\frac{[\lips(\chkdeg-1)(\vardeg-1)]^{\diam}}
     {\lips(\chkdeg-1)(\vardeg-1) - 1} \;\le\; \frac{\lbnd}{\vardeg},
\end{align*}
we have
\begin{align*}
\sum_{\fchk\in\neig(\svar)} |\avesdmes_{\fchksvar}^{\Time+1} -
\bpmes_{\fchksvar}^{\Time+1}| \; \le \; \lbnd, \quad \text{for all}
\quad \svar = 1, 2, \ldots, \numvar,
\end{align*}
and hence $\ffun{\svar}{\Time+1}{\avesdmes} +
\gfun{\svar}{\Time+1}{\avesdmes}\ge \lbnd$, which proves the
claim~\eqref{EqnClaim} and concludes the lemma.


\section{Proof of Lemma~\ref{LemNilpotent}}
\label{AppProofNilpotent}

We first show, via induction, that for any positive integer $\ell$ and
edges $(\fchkfvar)$, and $(\schksvar)$, the entry
$\mat^{\ell}(\fchkfvar, \schksvar) \neq 0$ if and only if there exists
a directed path of length\footnote{Here the length of the path is
  equal to the number of intermediate variable nodes plus one.} $\ell$
between the edges $(\schksvar)$, and $(\fchkfvar)$. More specifically,
there must exists a sequence of non-overlapping, directed
(check-variable) edges $\{(\genedge{\fchk_1}{\fvar_1}), \ldots,
(\genedge{\fchk_{\ell-1}}{\fvar_{\ell-1}})\}$ such that
\begin{align*}
\schk\in\neig(\svar)\setminus\{\fchk_{\ell-1}\}, \; 
\svar\in\neig(\fchk_{\ell-1})\setminus\{\fvar_{\ell-1}\}, \; \ldots, \; 
\fchk_1\in\neig(\fvar_1)\setminus\{\fchk\}, \;
\fvar_1\in\neig(\fchk)\setminus\{\fvar\}.
\end{align*}
The base case for $\ell = 1$, is obvious from
construction~\eqref{EqnDefnMat}. Suppose the claim is correct for
$\ell$; the goal is to prove it for $\ell+1$. By definition, we have
\begin{align*}
\mat^{\ell+1}(\fchkfvar, \schksvar) \; = \; \sum_{(\genedge{c}{k})}
\mat^{\ell}(\genedge{\fchk}{\fvar}, \genedge{c}{k}) \:
\mat(\genedge{c}{k}, \genedge{\schk}{\svar}).
\end{align*}
Since the matrix is non-negative, $\mat^{\ell+1}(\fchkfvar, \schksvar)
\neq 0$, if and only if there exists an edge $(\genedge{c}{k})$ such
that $\mat^{\ell}(\genedge{\fchk}{\fvar}, \genedge{c}{k}) \neq 0$, and
$\mat(\genedge{c}{k}, \genedge{\schk}{\svar}) \neq 0$. Therefore,
according to the induction hypothesis, there exist a sequence of
non-overlapping edges \mbox{$\{(\genedge{\fchk_1}{\fvar_1}), \ldots,
(\genedge{\fchk_{\ell-1}}{\fvar_{\ell-1}})\}$} such that
\begin{align*}
c\in\neig(k)\setminus\{\fchk_{\ell-1}\}, \; 
k\in\neig(\fchk_{\ell-1})\setminus\{\fvar_{\ell-1}\}, \; \ldots, \; 
\fchk_1\in\neig(\fvar_1)\setminus\{\fchk\}, \;
\fvar_1\in\neig(\fchk)\setminus\{\fvar\}.
\end{align*}
Moreover, we should have $\schk\in\neig(\svar)\setminus\{c\}$, and
$\svar\in\neig(c)\setminus\{k\}$. Putting the pieces together, there
must exists a directed path of length $\ell+1$, consisting of
\mbox{$\{(\genedge{\fchk_1}{\fvar_1}), \ldots,
  (\genedge{\fchk_{\ell-1}}{\fvar_{\ell-1}}), (\genedge{c}{k})\}$},
between edges $(\schksvar)$, and $(\fchkfvar)$, which yields the
claim. Finally, since there is no directed path longer than $\diam$
(diameter) between any pair of edges in a tree-structured factor
graph, we must have $\mat^{\diam} = 0$ that concludes the proof.


\section{Proof of Theorem~\ref{ThmMain}}
\label{AppProofThmMain}

As stated previously, proof of Theorem~\ref{ThmMain} is similar to
that of Theorem~\ref{ThmTree}. The major difference lie in the fact
that due to its cycle-free structure, tree-respecting matrix $\mat$,
defined in~\eqref{EqnDefnMat}, is nilpotent (recall the result of
Lemma~\ref{LemNilpotent}). However, the same may not necessarily be
true for general graphs. As a consequence, for a non-tree factor
graph, the right hand side of the inequality~\eqref{EqnUnwrapped} has
indeed a growing number of terms as $\Time\to\infty$. However, we can
upper-bound the error over a finite horizon provided by the stopping
time $\ftime$, defined in~\eqref{EqnDefnStop}. More precisely,
unwrapping the recursion~\eqref{EqnVectorRecursion} for $\tau = 0, 1,
\ldots, \Time$, taking norm-infinity on both sides of the outcome, and
doing some algebra yields
\begin{align*}
\max_{(\fchkfvar)} |\avesdmes_{\fchkfvar}^{\Time+1} -
\bpmes_{\fchkfvar}^{\Time+1}| \;& \le\; \frac{(\chkdeg-1)
  (1-\lbnd)^{\dimn+1}}{\dimn \: \lbnd} \:
\frac{[\lips(\chkdeg-1)(\vardeg-1)]^{\Time+1}}
     {\lips(\chkdeg-1)(\vardeg-1) - 1},
\end{align*}
for all $\Time = 0, 1, \ldots, \ftime-1$. Therefore, for $\dimn$
sufficiently large, specifically when 
\begin{align*}
\frac{(\chkdeg-1) (1-\lbnd)^{\dimn+1}}{\dimn \: \lbnd} \:
\frac{[\lips(\chkdeg-1)(\vardeg-1)]^{\ftime}}
     {\lips(\chkdeg-1)(\vardeg-1) - 1} \;\le\; \frac{\lbnd}{\vardeg},
\end{align*}
we have
\begin{align*}
\sum_{\fchk\in\neig(\svar)} |\avesdmes_{\fchksvar}^{\Time+1} -
\bpmes_{\fchksvar}^{\Time+1}| \; \le \; \lbnd,
\end{align*}
for all $\svar = 1, 2, \ldots, \numvar$, and $\Time = 0, 1, \ldots,
\ftime-1$, thereby completing the proof of a slightly different
version of the Lemma~\ref{LemMessBnd} over a finite horizon.

\begin{lemma}\label{LemMessBndGeneral}
For sufficiently large $\dimn$, there exists a fixed positive constant
\mbox{$\lbnd = \lbnd(\ftime) < 1$} such that
\begin{align*}
\min \big\{\ffun{\fvar}{\Time}{\bpmes} + \gfun{\fvar}{\Time}{\bpmes},
\ffun{\fvar}{\Time}{\avesdmes} + \gfun{\fvar}{\Time}{\avesdmes}\big\} \; \ge
\; \lbnd,
\end{align*}
for all $\Time = 0, 1, \ldots, \ftime$, and $\fvar = 1, 2, \ldots,
\numvar$. Furthermore, we have
\begin{align}\label{EqnMessBndGeneral}
\max_{(\fchkfvar)} |\avesdmes_{\fchkfvar}^{\Time} -
\bpmes_{\fchkfvar}^{\Time}| \;\le\; \frac{(\chkdeg-1)
  (1-\lbnd)^{\dimn+1}}{\dimn \: \lbnd} \:
\frac{[\lips(\chkdeg-1)(\vardeg-1)]^{\Time}}
     {\lips(\chkdeg-1)(\vardeg-1) - 1},
\end{align}
for all $\Time= 0, 1, \ldots, \ftime$.
\end{lemma}

\noindent Now substituting the inequality~\eqref{EqnMessBndGeneral}
into~\eqref{EqnMargError}, setting
\begin{align*}
\dimn \;=\; \max\left\{\frac{\log{\accu} - \ftime\log(\lips
  (\chkdeg-1)(\vardeg-1))}{\log(1 - \lbnd)} \, , \, \frac{3}{\lbnd}
\right\},
\end{align*}
we obtain
\begin{align*}
\max_{0\le\fvar\le\numvar} |\Expt\big[\sdmarg_{\fvar}^{\Time}\big] -
\bpmarg_{\fvar}^{\Time}|\; \le \; \accu,
\end{align*}
for all $\Time=0,1,\ldots,\ftime$, which concludes proof of part (a)
of Theorem~\ref{ThmMain}. Proof of part (b) follows the exact same
steps outlined in the Section~\ref{SubSecPartB}


\bibliographystyle{plain}
\bibliography{SD_bibfile}

\end{document}